\documentclass[aps,pre,twocolumn,amsmath,amssymb,superscriptaddress,floatfix]{revtex4}

\usepackage{graphicx}
\usepackage{bm}
\usepackage[usenames]{color}
\bibstyle{apsrev.bib}

\newcommand{\be}{\begin{equation}}
\newcommand{\ee}{\end{equation}}
\newcommand{\beqn}{\begin{eqnarray}}
\newcommand{\eeqn}{\end{eqnarray}}

\begin{document}

\title{Long-range epidemic spreading in a random environment}
\author{R\'obert Juh\'asz}
\email{juhasz.robert@wigner.mta.hu}
\affiliation{Wigner Research Centre for Physics, Institute for Solid State
Physics and Optics, H-1525 Budapest, P.O.Box 49, Hungary}
\author{Istv\'an A. Kov\'acs}
\email{kovacs.istvan@wigner.mta.hu}
\affiliation{Wigner Research Centre for Physics, Institute for Solid State
Physics and Optics, H-1525 Budapest, P.O.Box 49, Hungary}
\affiliation{Institute of Theoretical Physics, Szeged University, H-6720 Szeged,
Hungary}
\affiliation{Center for Complex Networks Research and Department of Physics,
Northeastern University, 110 Forsyth Street, 111 Dana Research Center,
Boston, MA 02115, USA}
\author{Ferenc Igl\'oi}
\email{igloi.ferenc@wigner.mta.hu}
\affiliation{Wigner Research Centre for Physics, Institute for Solid State
Physics and Optics, H-1525 Budapest, P.O.Box 49, Hungary}
\affiliation{Institute of Theoretical Physics, Szeged University, H-6720 Szeged,
Hungary}
\date{\today}

\begin{abstract}
Modeling long-range epidemic spreading in a random environment, we consider a quenched disordered, $d$-dimensional contact process with infection rates decaying with the distance as $1/r^{d+\sigma}$.  
We study the dynamical behavior of the model at and below the epidemic threshold by a variant of the strong-disorder renormalization group method and
by Monte Carlo simulations in one and two spatial dimensions. 
Starting from a single infected site, the average survival probability is found to decay as $P(t) \sim t^{-d/z}$ up to multiplicative logarithmic corrections.
Below the epidemic threshold, a Griffiths phase emerges, where the dynamical exponent $z$ varies continuously with the control parameter and tends to $z_c=d+\sigma$ as the threshold is approached. 
At the threshold,  the spatial extension of the infected cluster (in surviving trials) is found to grow as $R(t) \sim t^{1/z_c}$ with a multiplicative logarithmic correction, and the average number of infected sites in surviving trials is found to increase as $N_s(t) \sim (\ln t)^{\chi}$ with $\chi=2$ in one dimension. 
\end{abstract}


\maketitle

\section{Introduction}

The contact process (CP) \cite{cp,liggett} is a basic model in the fields of 
epidemic spreading and population dynamics. It is
defined on a lattice, the sites of which are either active or
inactive. The activity spreads to other inactive sites (which is
interpreted as spreading of the infection in case of epidemics
modeling) or vanishes (which is interpreted as the spontaneous
recovery of the individual at that site) stochastically with given
rates.  This model has attracted much interest because it undergoes a
continuous phase transition from a fluctuating active phase to an
inactive one if the relative magnitude of infection rates with respect
to the recovery rate is varied.  In its simplest form, activity
spreads to neighboring lattice sites with uniform rates. The phase
transition of this variant falls into the universality class of
directed percolation (DP) and the critical
exponents are known with a high precision
\cite{md,odor,hhl}, although their exact values are unknown.

The contact process was later generalized in different directions and studied
with a focus on the phase transition.  
Here, we highlight two factors
that have a substantial effect on the critical behavior.  In the
long-range contact process (LRCP), infection is not restricted to
neighboring sites but allowed to occur on any other site with
a rate that decays as a power of the distance $\lambda(r)\sim
r^{-(d+\sigma)}$, where $d$ denotes the dimension \cite{mollison}.
The critical exponents of this model have been calculated to first
order in an $\epsilon$-expansion by a field-theoretic renormalization
group method \cite{janssen}. It turned out that, in analogy with the long-range
Ising and $O(N)$ models\cite{fisher_me,sak,bloete,picco,parisi}, the critical exponents
vary continuously with $\sigma$ in the range $\sigma_{\rm
  MF}(d)<\sigma<\sigma_{\rm DP}(d)$, while, for $\sigma>\sigma_{\rm
  DP}(d)$, the model is in the DP universality class, and for
$\sigma<\sigma_{\rm MF}(d)$, the critical behavior is described by a
mean-field theory. The upper critical dimension of the LRCP is $d_c={\rm
min}(4,2 \sigma)$
and the correlation length critical exponents in the mean-field region ($d>d_c$)
are $\nu^{\perp}_{MF}={\rm max}(1/2,1/\sigma)$
and $\nu^{\parallel}_{MF}=1$. In the intermediate region $\nu^{\perp}(\sigma)$ ($\nu^{\parallel}(\sigma)$)
is a monotonically decreasing (increasing) function of $\sigma$. Numerical simulations in one dimension have been
confirmed the above scenario \cite{howard}.  Motivated by the dynamics
of a fluctuating interface growing on a one-dimensional substrate,
a restricted variant of the LRCP has been studied, as well
\cite{ginelli,fiore}. In this model, activation of a site occurs only
by the nearest active site. The main difference in critical behavior of this variant compared to that of the unrestricted one is the absence of a mean-field region. 
Long-range epidemic spreading has also been studied by other models, such as 
the susceptible-infected-recovered model \cite{linder,grassberger}.  
For further studies in this field, we mention Refs. \cite{ginelli1,adamek,hinrichsen}.

Another circumstance that changes the critical behavior of CP is 
{\it quenched disorder}. 
According to a strong-disorder renormalization
group (SDRG) study in one dimension \cite{hiv}, the critical behavior
is, at least for strong enough initial disorder, described by a so
called {\it infinite-disorder fixed point} (IDFP)\cite{fisher,im}, where the
relationship between time, $\tau$, and length scale, $\xi$, is extremely
anisotropic, $\ln\tau\sim \xi^{\psi}$. As a consequence, the disorder-averaged
dynamical quantities such as the survival probability decay as a power
of $\ln t$ rather than as a power of $t$. Furthermore, in a close
analogy to the Griffiths phase of quantum magnets \cite{griffiths},
the dynamics in the inactive phase is characterized by power laws,
where the exponents vary continuously with the control parameter
\cite{noest,vojta_rev}.

In this work, we aim at studying the dynamics of the CP in the
simultaneous presence of the above components, i.e. long-range
interactions and quenched disorder. This type of problem emerges
in different situations. Here we mention the case, when an infectious disease
is transmitted by insects, which make occasionally long flights, the
length of which has a power-law distribution. Similarly, human infections can
be transported through long-distance plane flights or plant disease
epidemics is spread through atmospheric dispersal, which has also a power-law
characteristics \cite{reynolds}. In all cases the infection and recovery rates are generally position dependent random
variables. We shall see that the interplay of disorder and long-range
interactions results in a critical behavior, which is different from those
observed in the presence of only one component.  The long-range random
transverse-field Ising chain, which is closely related to present
problem due to a formally similar SDRG treatment, has been recently
studied with the conclusion that the critical behavior is controlled
by a strong-disorder fixed point with a power law scaling \cite{jki}.
By inspecting the SDRG scheme of the CP in one dimension, we will show
that, in spite of the differences in the SDRG rules, the same
conclusions hold also for this model. In addition to this, we will present
results of a numerical SDRG study in two dimensions, as well as
results of Monte Carlo simulations in one and two dimensions.

The rest of the paper is organized as follows. The model is defined in
Sec.  \ref{model} and a phenomenological argument about a limiting
value of the dynamical exponent in the Griffiths phase is given in
Sec. \ref{phenomenology}. A SDRG scheme of the model is presented
in Sec. \ref{sdrg}. Our results for the one-dimensional model are 
presented in Sec. \ref{1d} and for the two-dimensional model 
in Sec. \ref{2d}. Finally, the results are discussed in
Sec. \ref{discussion} and the details of the SDRG calculations are
deferred to the Appendix.

\section{The model}
\label{model} 

Let us consider a d-dimensional cubic lattice, the sites of which can
be either active or inactive, and consider a continuous-time Markov
process with the following (independent) transitions. Site $i$, if it
is active, becomes spontaneously inactive with a rate $\mu_i$ or it
activates site $j$, provided the latter is inactive, with a rate
$\lambda_{ij}$.  The activation rates are parameterized as follows:
\be 
\lambda_{ij}=\Lambda_{ij}r_{ij}^{-(d+\sigma)},
\label{lambda}
\ee
where $r_{ij}$ is the Euclidean distance between site $i$ and $j$,
and $\Lambda_{ij}$ are $O(1)$ i.i.d. quenched random variables drawn
from some distribution $\pi(\Lambda)$. The recovery rates $\mu_i$ are
also i.i.d. quenched random variables drawn from a distribution
$\kappa(\mu)$.  
For the sake of brevity, we introduce the decay exponent $\alpha$ of the infection rates  as $\alpha\equiv d+\sigma$. 
We will restrict ourselves to the regime $\alpha>d$,
where the total rate of activation events from a given site,
$\lambda_i=\sum_{j\neq i}\lambda_{ij}$, remains finite.

This system exhibits two different phases. For a low enough 
tendency for recovery, such that 
$\overline{\ln \mu}-\overline{\ln \Lambda}=\theta < \theta_c$ the fraction
of active sites in the stationary state is $\rho >0$, 
which represents the active phase. (Here, and in the following, 
the overbar denotes an average over quenched disorder.) On the contrary, for a high enough tendency for recovery, $\theta > \theta_c$, we have $\rho = 0$ in the stationary state. 
In between, at
$\theta = \theta_c$, there is a non-equilibrium phase transition in the system,
the properties of which are the subject of this work.
  
Although we will consider a finite strength of disorder, as a first
step, it is worth investigating the stability of the fixed point of the clean system
(with long-range infections) against a weak disorder. Generalizing the heuristic criterion by
Harris \cite{harris,noest} for long-range interactions, weak disorder is predicted to be relevant if
\be 
d\nu^{\perp}<2,
\label{criterion}
\ee
where $\nu^{\perp}$ is the correlation-length exponent of the non-random
system. According to the known results for $\nu^{\perp}$, discussed in the Introduction, weak disorder is
generally relevant, except the mean-field region, where it is irrelevant.

\section{Phenomenological theory in the Griffiths phase}
\label{phenomenology}

In the short-range random CP, where activation occurs only on neighboring
sites, the average density, i.e. the fraction of active sites $\rho(t)$
decays as an inverse power of time starting from a fully active initial state, in a regime $\theta_c < \theta < \theta_G$ of the inactive phase, called as Griffiths phase \cite{noest}.
This type of semi-critical behavior (short-range
spatial, but long-range dynamical correlations) can be explained by
a phenomenological theory, for a review see
e.g. Ref. \cite{vojta_rev}.  According to this, although the entire
system is subcritical on average and tends toward the absorbing (inactive) state for long times, it still contains clusters of sites,
the so called {\it rare regions}, where the majority of internal
activation rates are greater than the average, so that the local control parameter, $\theta_l$,
is below the bulk critical point, thus these regions
are locally supercritical. A probability of occurrence of rare regions
is exponentially small but their lifetime is exponentially large in
their size, so that the distribution of their lifetimes has an
algebraic tail,
\be
P_>(\tau)\sim \tau^{-d/z},
\label{taudist}
\ee
characterized by a non-universal dynamical exponent $z$, which
depends on the distance from the
critical point: $z=z(\theta)$ and it diverges as
the critical point is approached.  If the initial state of the model
is the fully active one, then after a long time, the activity will
survive in clusters having a lifetime greater than $t$.  Let us denote
the characteristic distance between nearest active clusters at time
$t$ by $\ell(t)$. At time $t$, the fraction of active clusters is on
one hand in the order of $1/\ell^d(t)$, on the other hand it is
proportional to $P_>(t)$. Using Eq. (\ref{taudist}), we have thus
\be
\ell(t)\sim t^{1/z}\;,
\label{z}
\ee
the usual relation between length and time scales.
The typical (effective) activation rate between
nearest active clusters at time $t$ is exponentially small in
$\ell(t)$: $\lambda_{\rm eff}[\ell(t)] \sim \exp[-c\ell(t)]$,
so the active clusters, which have a typical lifetime $O(t)$
or, equivalently, $O[\ell^z(t)]$, do not interact with each other for
long times, and each of them arrives at the inactive state at a time
determined by its own lifetime, which has been tacitly assumed.  Using
that the mass of clusters (i.e. number of sites contained in it) that
are active at time $t$ is at most $m(t)=O(\ln t)$, one obtains for the
time-dependence of the average density 
\be 
\rho(t)\sim m(t)P_>(t)\sim t^{-d/z}
\label{rhot}
\ee
up to logarithmic corrections.

Let us now see how the above picture is modified in the case of long-range
interactions. In the following reasoning, we assume that the relevant time scale
in the problem is governed by the supercritical rare regions as in the short-range case
and thus the relations in Eqs.(\ref{taudist}) and (\ref{z}) are valid with an appropriate value of
the dynamical exponent, $z(\theta)$. But, due to the long-range interactions, the typical
activation rates between neighboring active clusters are different and these are estimated
to be $\lambda_{\rm eff}[\ell(t)]\sim \ell^{-\alpha}(t)m^2(t)$. 
Here, the typical mass of an active cluster $m(t)$ is expected to be a
slower-than-algebraic (logarithmic) function of the time.  
For a fixed pair of
neighboring active clusters, the characteristic time between two
subsequent activation events thus scales as $\tau_a\sim
1/\lambda_{\rm eff}[\ell(t)]\sim \ell^{\alpha}(t)m^{-2}(t)$. These clusters can be
regarded as independent if $\tau_a$ is much greater than the typical
lifetimes of the clusters, $\tau_l\sim \ell^z(t)$.  Independence is
realized asymptotically if $\tau_l(\ell)/\tau_a(\ell)\to 0$ as
$\ell\to\infty$ ($t\to\infty$), yielding 
\be
z<\alpha,
\ee
which is a
necessary condition for the self-consistency of the above
phenomenological picture.  We have thus obtained that, for a fixed
$\alpha$, the dynamical exponent in the Griffiths phase cannot
increase unboundedly as the critical point is approached, but this
phase must terminate at a point, where $z$ reaches the boundary value
$\alpha$.  At this point, the above picture breaks down since the
clusters no longer die out independently and the infection rates and
recovery rates become comparable with each other. This suggests that
this point coincides with the phase transition point separating the
Griffiths phase from the active phase, so that
\be
\lim_{\Theta\to\Theta_c+}z(\Theta)=\alpha.
\label{z_c}
\ee

\section{The strong-disorder renormalization group treatment}
\label{sdrg}

In the strong-disorder renormalization group approach to the contact
process \cite{hiv,im}, the transition with the highest rate is
iteratively eliminated and the effective rates of the reduced system
are calculated perturbatively. Thereby the actually highest rate
$\Omega$ of the renormalized system is gradually reduced. The
procedure consists of two kinds of reduction steps and here
we recapitulate the results in\cite{hiv}.  If the largest
rate is an activation rate, $\Omega=\lambda_{ij}$, site $i$ and $j$
form a cluster, which has an effective deactivation rate 
\be
\tilde\mu_{ij}\simeq\frac{2\mu_i\mu_j}{\Omega}
\label{mu_rule}
\ee
if $\mu_i,\mu_j\ll\lambda_{ij}$.  If the largest rate is a
deactivation rate, $\Omega=\mu_i$, cluster $i$ is deleted and clusters
that were neighboring to $i$ become directly connected through an
effective activation rate 
\be
\tilde\lambda_{jk}\simeq\frac{\lambda_{ji}\lambda_{ik}}{\Omega}
\ee
if
$\lambda_{ji}\lambda_{ik}\ll\mu_i$.  Apart from one-dimensional
systems, which the above procedure has been originally developed for
\cite{mdh,fisher}, double connections between clusters may appear
after performing the above reduction steps. This is treated in
practice in two different ways. The rates of the two parallel
transitions are either added, which is termed as {\it sum rule}, or
the larger one of them is kept, in which case we speak about {\it
  maximum rule}.  In case of an IDFP, which governs the critical behavior of the short-range CP, the
SDRG with both rules becomes asymptotically exact and is conjectured
to provide correct critical exponents \cite{im}.  The
random-transverse field Ising model (RTIM) has an SDRG scheme
identical to that of the CP, apart from the absence of factor $2$ in
Eq. (\ref{mu_rule}) \cite{fisher}. This difference is irrelevant in the
critical properties of the two models for short-range interactions and we
expect the same conclusion for the long-range models, too.

For the RTIM, the SDRG procedure with
the maximum rule can be implemented in a very efficient numerical algorithm\cite{kovacs},
which provides accurate results for the short-range model in higher dimensions, too.
This numerical procedure has been used to the one-dimensional long-range
RTIM, and the observed renormalization steps are summarized in an approximate
primary model, which has been analytically solved\cite{jki,stretched}. Here,
we generalize these steps for the one-dimensional LRCP and the results
are summarized in Sec.\ref{1d}. For the two-dimensional model, results
of the numerical analysis with the RTIM maximum rule are presented in Sec.\ref{2d}.

\section{Results for the one-dimensional model}
\label{1d}

\subsection{SDRG analysis}
\label{sdrg1d}
A simplified SDRG scheme of the one-dimensional random LRCP, which we call primary scheme, is presented in the Appendix together with its analytical solution. 
In this approach, the renormalized system has a one-dimensional structure with effective interactions only between neighboring (non-decimated) clusters. 
Long-range interactions between remote clusters are taken into account only at a later stadium of the renormalization procedure, when these clusters become nearest neighbors.  
A further simplification in the primary scheme is that  
the effective interactions between clusters are approximated by the long-range interaction between the closest constituent sites of clusters. 
In this respect, this treatment is closer to a disordered variant of the restricted LRCP mentioned in the Introduction \cite{ginelli,fiore}.
Unfortunately, an extension of the primary scheme, which takes into account all pair interactions between neighboring clusters and which is more appropriate to describe the unrestricted model cannot be solved {\it ab initio}; nevertheless, its scaling properties can be inferred from those of the primary scheme by a heuristic reasoning. 

Before presenting the results obtained by the SDRG method, a caveat is in order concerning its reliability. We will see, that the critical behavior of the model is described by a strong-disorder fixed-point, where the disorder remains finite, rather than by an IDFP. In this case, the asymptotic exactness of the method is not guaranteed, therefore the outcomes of the method must be confronted with results of Monte Carlo simulations of the original model. 
Nevertheless, for another example of a model with a strong-disorder fixed point, the two-dimensional, random Heisenberg model namely, the predictions of the method have been proven qualitatively correct \cite{lin}.

The main results of the analysis of the primary scheme are summarized in Eqs. (\ref{dyn00}) and (\ref{Qell}) below.
A relationship between the rate scale $\Omega$ and the length scale
$\ell$, which is the inverse of the concentration of non-decimated
clusters is provided by the method in the form
\be
\ell\simeq\ell_0\left(\frac{\Omega_0}{\Omega}\right)^{1/\alpha}\left[\frac{1}{
\alpha}\ln\frac{\Omega_0}{\Omega}\right]^2
\label{dyn00}
\ee
which follows from Eq.(\ref{dyn0}), where $\Omega_0$ is the initial rate scale
and $\ell_0$ is a non-universal constant depending on the distribution of rates.
Another important quantity is the survival probability $Q(\Omega)$ of
a given site during the SDRG procedure, which is found to decay with
$\Omega$ as given in Eq. (\ref{Qgamma}), or in terms of the length
scale, 
\be 
Q(\ell)\sim
\frac{\ell_0}{\ell}\left[\ln\frac{\ell}{\ell_0}\right]^2.
\label{Qell}
\ee
The results in Eqs. (\ref{dyn00}) and (\ref{Qell}) are obtained for the primary
model, in which the SDRG scheme is based on the maximum rule. These results
can be somewhat improved by using a scheme, which works partially by the sum rule, see
in\cite{jki} for the RTIM. It
takes, namely, in account the interactions between all pairs of sites
of neighboring clusters and, as it has been argued and confirmed
numerically, this improved scheme is roughly mapped to the primary scheme
if it is formulated in terms of reduced variables, $\lambda/m^2(\ell)$
and $\mu/m^2(\ell)$, where $m(\ell)=Q(\ell)\ell$ is the average mass
of non-decimated clusters at the length scale $\ell$.  This means that
improved scaling relations are obtained from those of the primary
scheme by replacing $\Omega$ with $\Omega/[Q(\ell)\ell]^2$. This leaves
the scaling of $Q(\ell)$ in Eq. (\ref{Qell}) in leading order
unchanged, but affects the power of the logarithmic correction in the
dynamical relation in Eq. (\ref{dyn00}), resulting in 
\be
\ell\sim
\left(\frac{\Omega_0}{\Omega}\right)^{1/\alpha}
\left(\ln\frac{\Omega_0}{\Omega}\right)^{2+4/\alpha}.
\label{dyn}
\ee
We will use this relation in the following.

\subsection{Scaling at criticality}
The scaling properties of different observables at the critical point
can be obtained from Eqs. (\ref{Qell}) and (\ref{dyn}) as follows.
Starting the process from a fully active state, Eq. (\ref{dyn})
provides a relationship for the typical time $t=\Omega^{-1}$ needed
for a segment of size $\ell$ in an infinite system or, for a finite
system of size $\ell$, to settle in a quasi-stationary state, in which
only the largest cluster identified by the SDRG method in the given
segment is active and the other sites are typically inactive.  The
average fraction of active sites $\rho(L)$ in the quasi-stationary
state of finite systems of size $L$ scales as    
\be 
\rho(L)\sim Q(L)\sim
L^{-1}(\ln L)^2,
\label{rhoL}
\ee
where we have used Eq. (\ref{Qell}).  However, in the numerical
simulations to be presented in the next section, we have studied the
time-dependence of various quantities as the process
approaches the stationary state rather than considering stationary ones.
To be concrete, starting the process from a single active site, we are
interested in the time-dependent average survival probability $P(t)$,
which is the probability that the process has not yet trapped in the
absorbing (inactive) state at time $t$. Another quantities
commonly studied with the above setup is the average number $N(t)$ of
active sites at time $t$ and the spread characterizing the spatial
extension of the growing cluster of active sites, defined as
   \be 
   R(t)=\exp\left\{\overline{\left\langle\frac{\sum_{i\neq
      0}n_i(t)\ln r_i}{\sum_{i}n_i(t)}\right\rangle_{s}}\right\}.
\label{spread}
\ee 
Here $n_i(t)$ is $1$ ($0$) if site $i$ at time $t$ is active
(inactive), $r_i$ denotes the Euclidean distance from the initially
active site $0$, $\langle\cdot\rangle_{s}$ denotes the expectation
value under the condition that the process is active at time $t$ in a
fixed random environment (i.e. set of transition rates).
Note that the
common definition of the spread for short-range CP through the second
moment of the distance from the origin would be divergent in the LRCP
for any finite $t$, hence the average of the logarithmic distance,
which is finite, is considered here instead \cite{howard}.  Due to the self-dual
property of the CP \cite{liggett,hv}, the average survival probability
$P(t)$ equals to the average density $\rho(t)$ in the case when the
process had been started from a fully active state.  Expressing the
length $L$ in Eq. (\ref{rhoL}) with time $t \sim 1/\Omega$ using Eq. (\ref{dyn}), we
obtain for the asymptotic time-dependence of the average survival
probability in the critical point 
\be
P(t)=\rho(t)\sim [t(\ln
  t)^4]^{-\frac{1}{\alpha}}.
\label{Pt}
\ee
We can see, that the critical decay exponent $1/\alpha$ coincides
with the limiting value of decay exponent at the upper boundary of the
Griffiths phase found by the phenomenological theory in the previous
section, see Eq. (\ref{z_c}).  
The average number of active sites scales with the length
$\ell$ as $N(\ell)\sim \frac{(\ln\ell)^2}{\ell}(\ln\ell)^2$, where the
first factor is the probability that the starting site was part of the
largest cluster in a segment of length $\ell$ and the second one is
the mass of the largest cluster.  Using Eq. (\ref{dyn}), we obtain    
\be
N(t)\sim [t(\ln t)^{4-2\alpha}]^{-\frac{1}{\alpha}}.  
\ee 
for large
$t$.  Thus, the average number $N_s(t)$ of active sites in surviving
samples scales with time as    
\be 
N_s(t)=\frac{N(t)}{P(t)}\sim (\ln t)^2
\label{Nst}
\ee 
irrespectively of $\alpha$.  As it has been argued in
Ref. \cite{jki}, the length of the longest bond in the largest cluster
of a segment of length $\ell$ is $O[\ell/(\ln\ell)^4]$.  Assuming that
the length of the cluster, which roughly determines the spread, is in
this order of magnitude and using Eq. (\ref{dyn}), the spread is
expected to scale with time asymptotically as 
\be
R(t) \sim [t(\ln
  t)^{4-2\alpha}]^{\frac{1}{\alpha}}.
\label{Rt}
\ee
\subsection{Numerical simulations}
\label{simulation1d}

We have performed discrete-time Monte Carlo simulations of the disordered
LRCP on rings of $L=10^9$ sites. The disorder was implemented by a
random dilution, i.e. the fraction $c$ of sites was removed. The
simulation consisted of the following moves. An active site is chosen
randomly and, with a probability $1/(1+\lambda)$, it is made inactive, or,
with a probability $\lambda/(1+\lambda)$, a random variable $r$ from an algebraic
distribution with the probability density $f(r)=(\alpha-1)r^{-\alpha}$
in the range $(1,\infty)$ is generated.  Then one of the two sites
whose distance from the active site is the integer part of $r\mod L$
is chosen with equal probabilities as a target site. If the target
site is an existing and inactive site, it will be activated.  One
Monte Carlo step (of unit time) consists of $n(t)$ such moves, where
$n(t)$ is the number of active sites at the beginning of the Monte
Carlo step.  Starting the process from a single active site, we have
followed the simulation up to $t=2^{27}$ Monte Carlo steps. Repeating
the simulation for $10-1000$ randomly diluted lattices and for $10^5$
starting positions per sample, we have calculated the average survival
probability $P(t)$, the average number of active sites $N(t)$ and the
spread $R(t)$ as a function of time for different values of the decay
exponent $\alpha$, the dilution parameter $c$, and the control
parameter $\lambda$.

\begin{figure*}
\includegraphics[width=5.6cm]{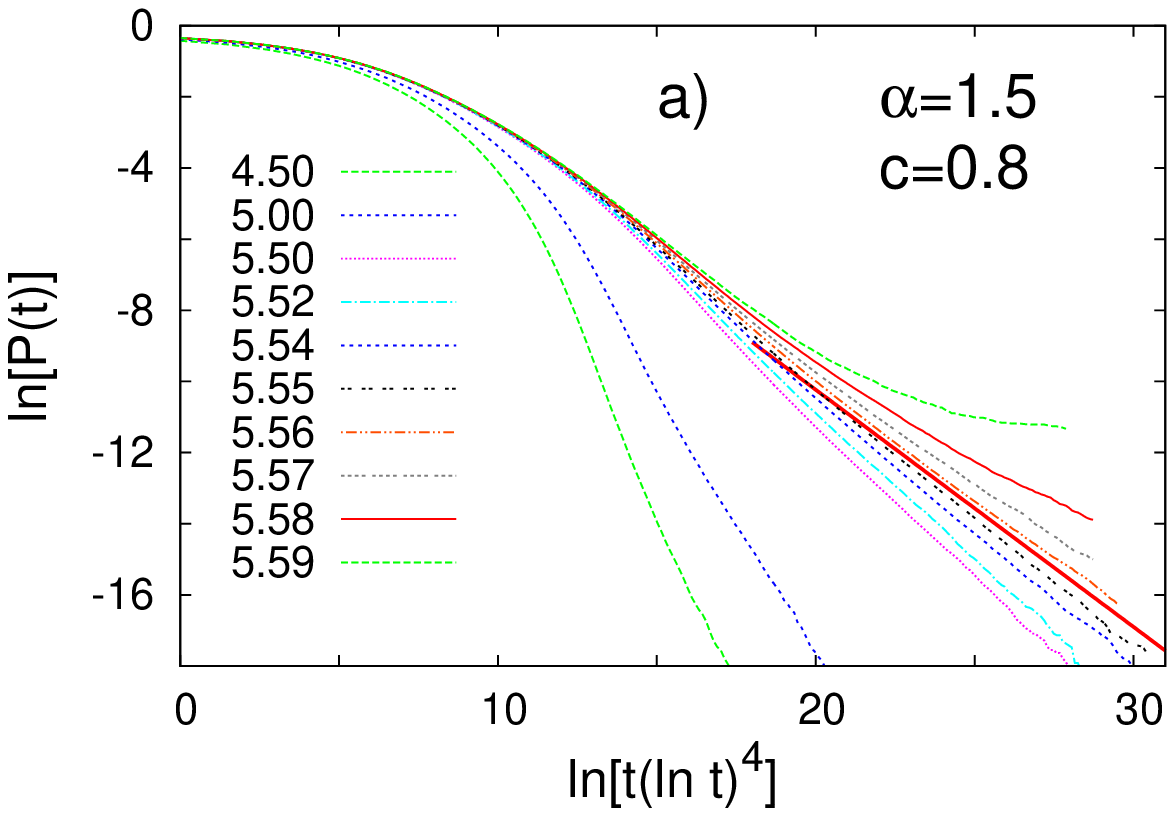}   
\includegraphics[width=5.6cm]{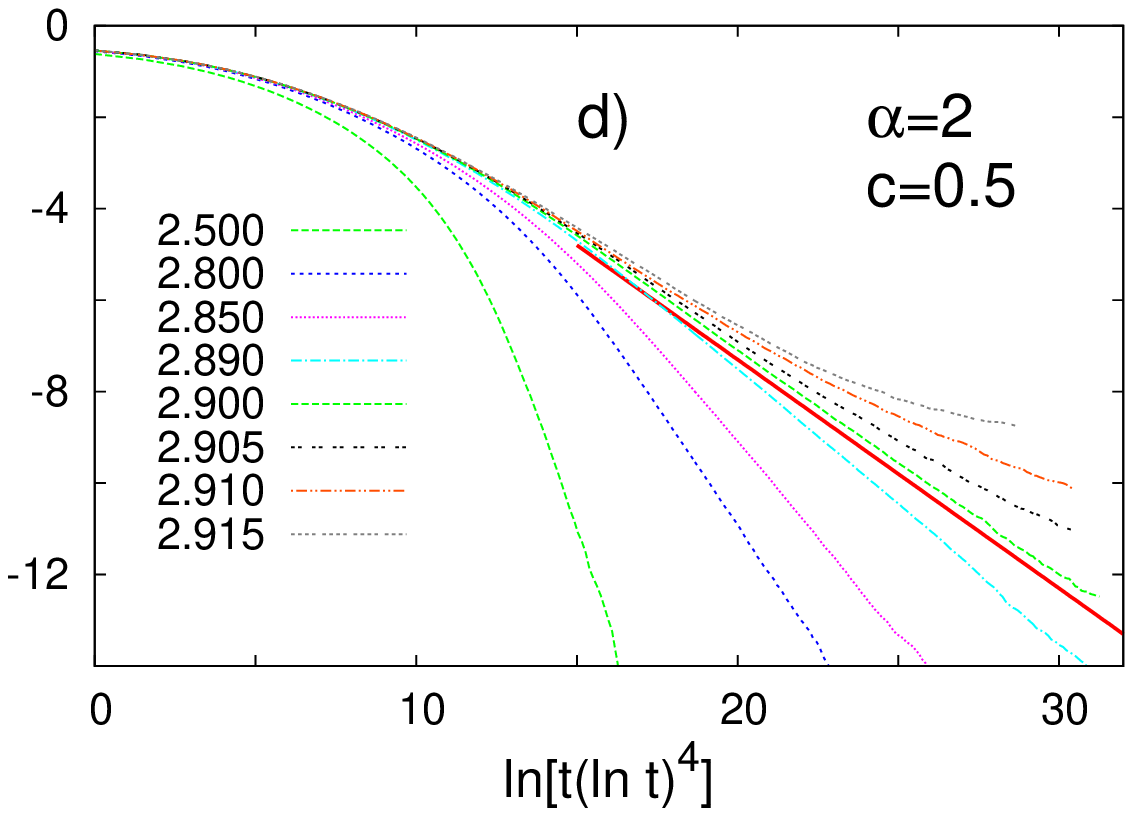}   
\includegraphics[width=5.6cm]{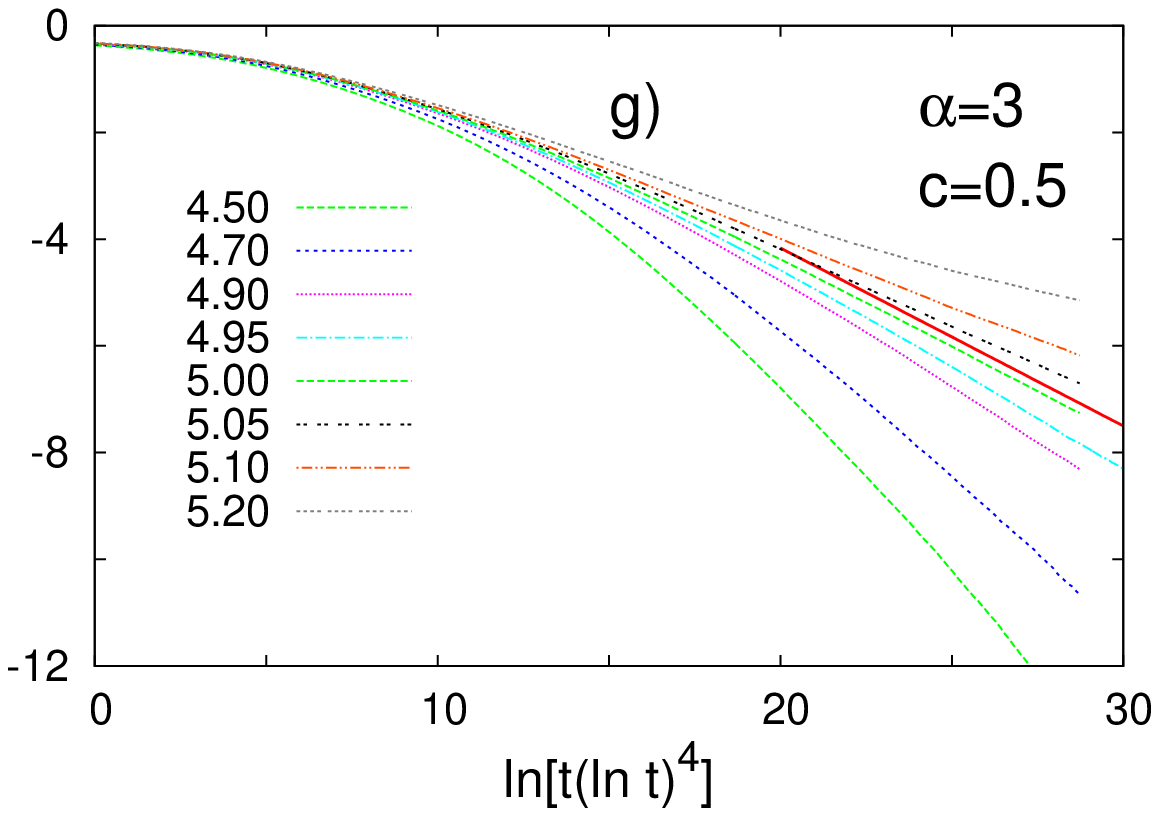}   
\includegraphics[width=5.6cm]{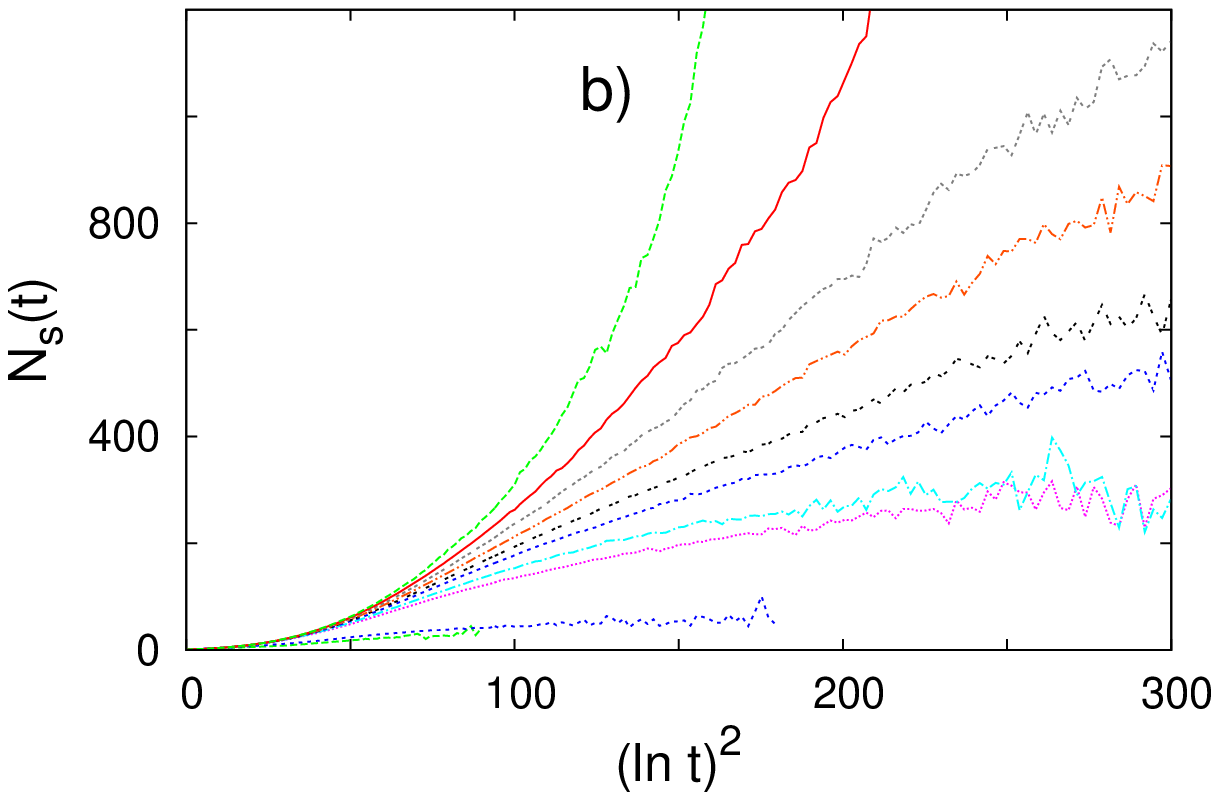}   
\includegraphics[width=5.6cm]{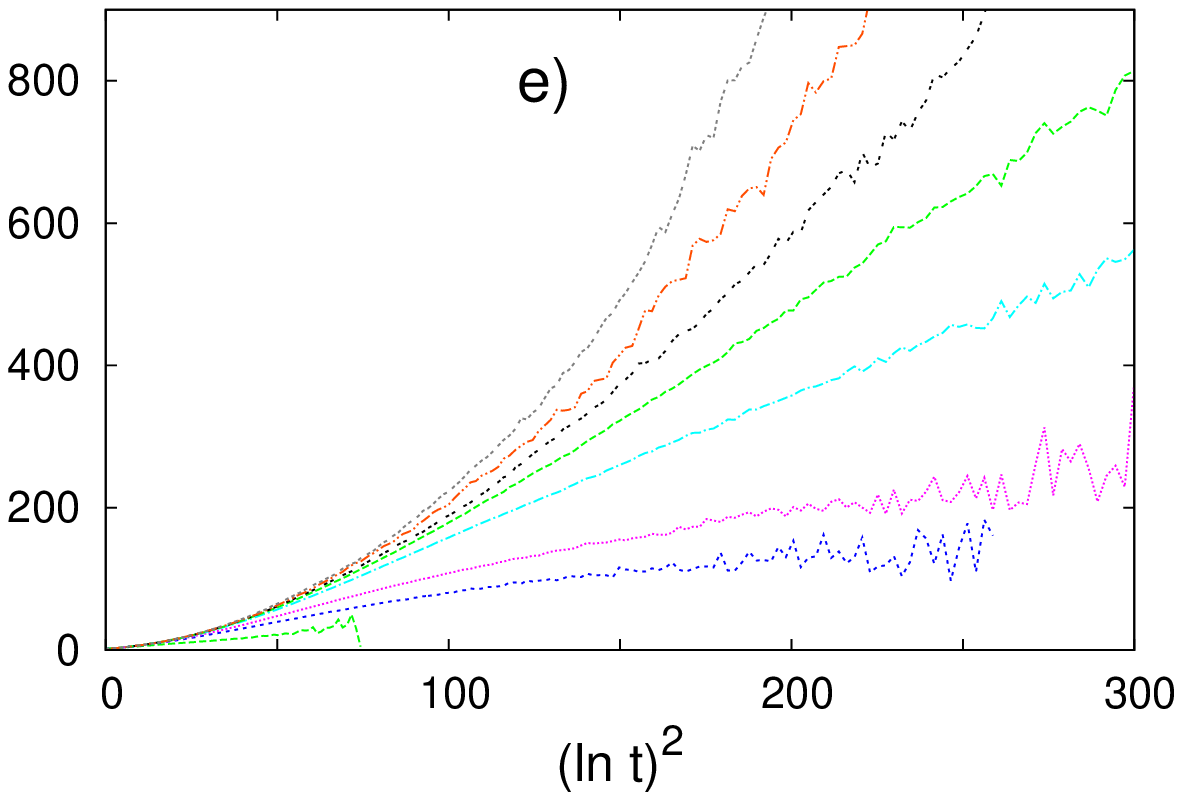}   
\includegraphics[width=5.6cm]{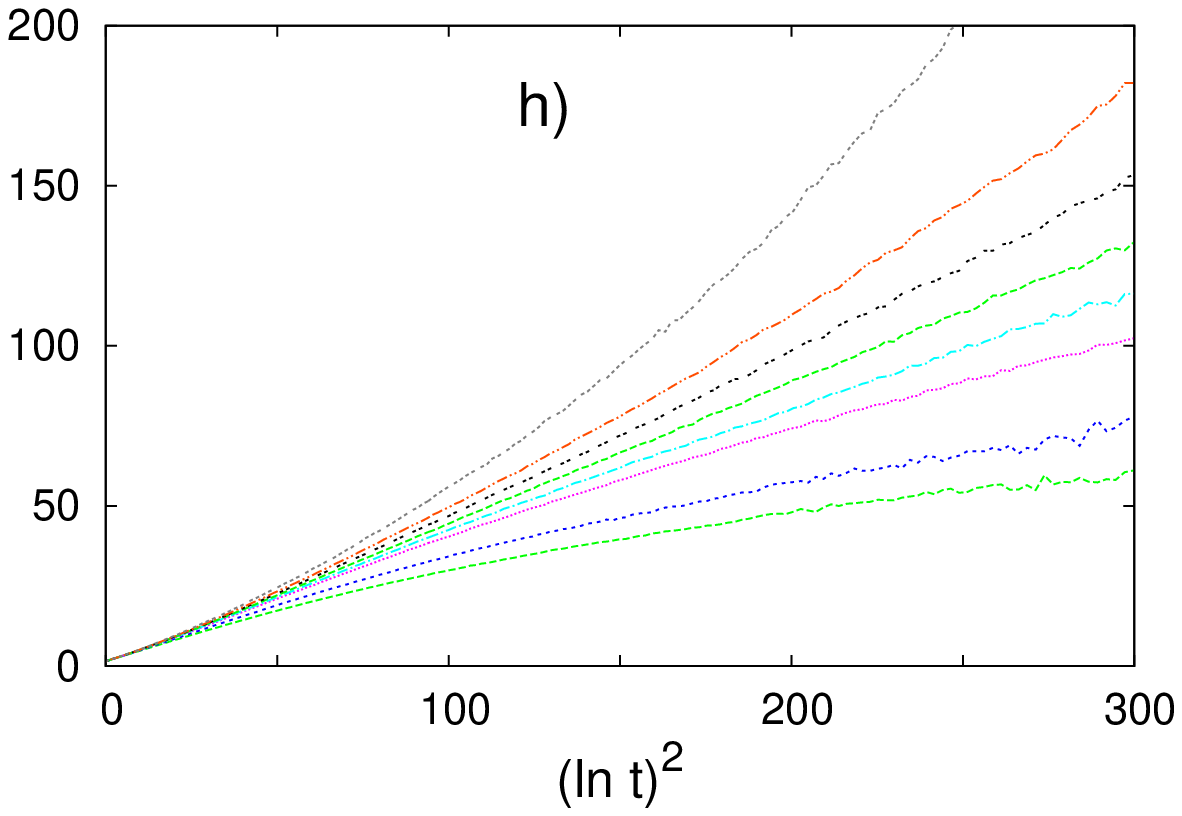}   
\includegraphics[width=5.6cm]{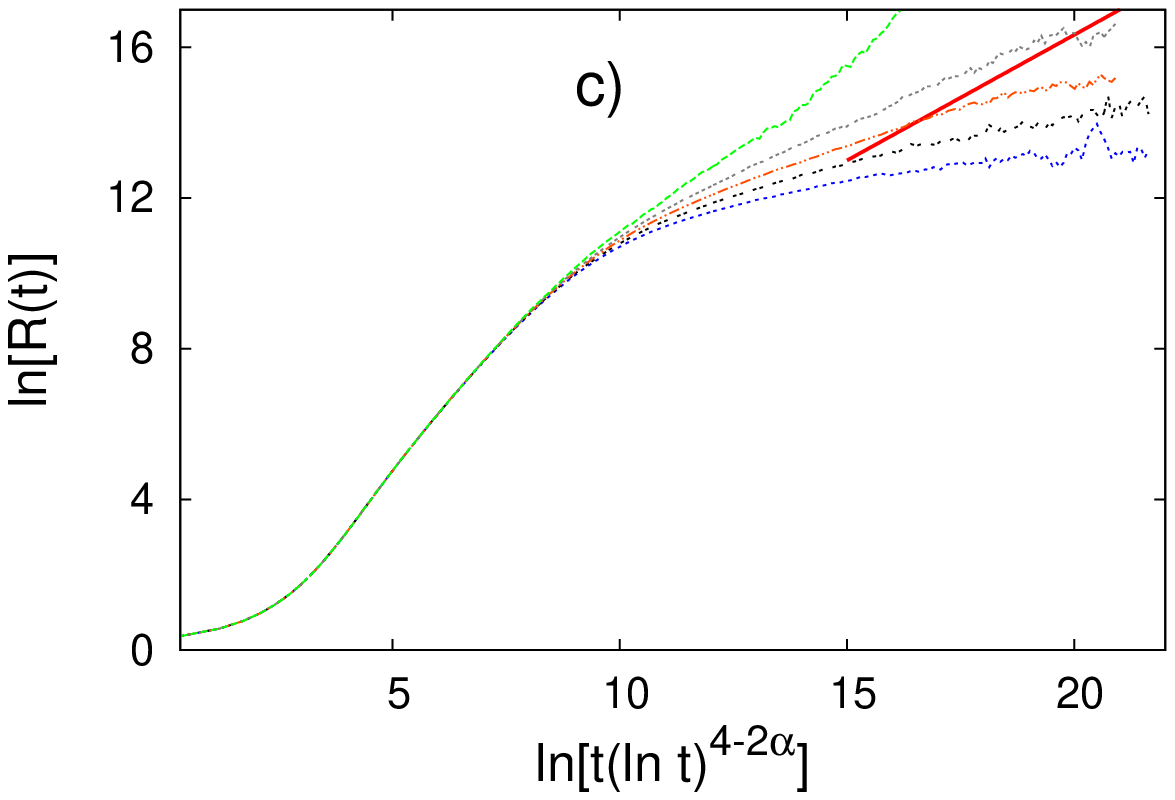} 
\includegraphics[width=5.6cm]{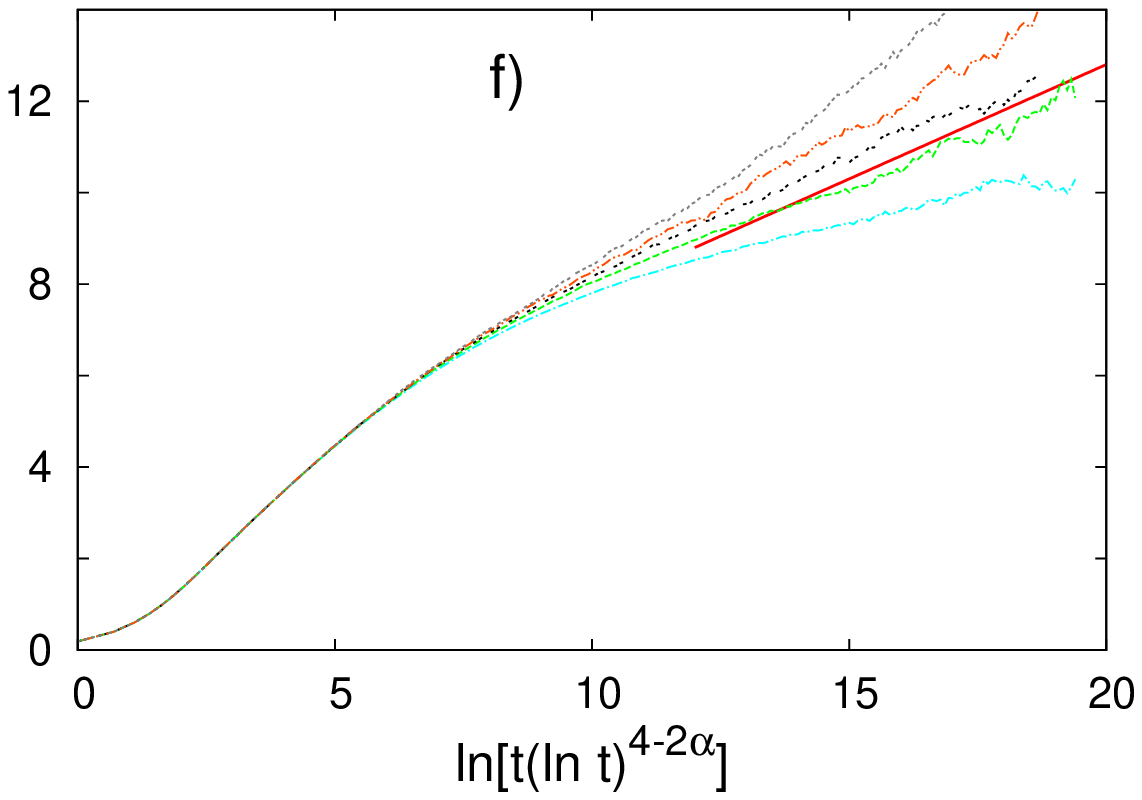}   
\includegraphics[width=5.6cm]{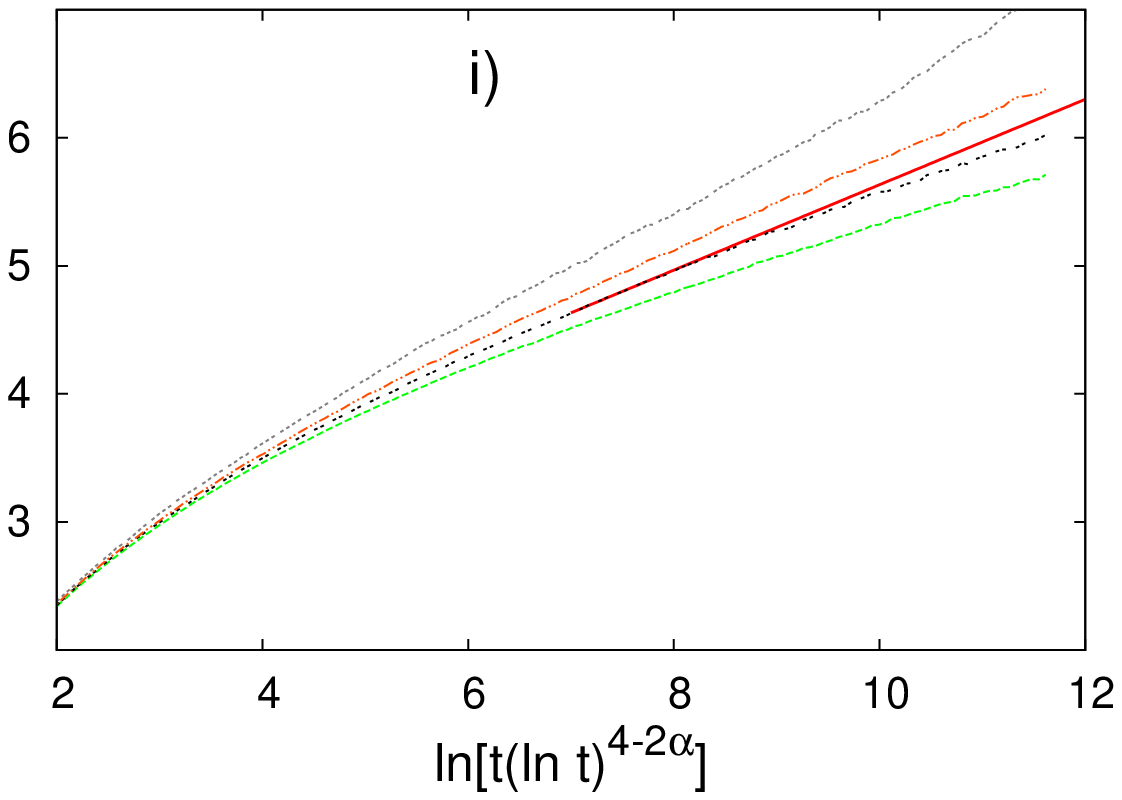}   
\caption{\label{1dfigs} (Color online) First row: The logarithm of the
  average survival probability $P(t)$ plotted against $\ln[t(\ln
    t)^4]$.  The data has been obtained by numerical
  simulations of the one-dimensional model for different values of the control
  parameter $\lambda$.  The solid line with a slope $-1/\alpha$
  indicates the asymptotic behavior predicted by the SDRG approach of
  the model in the critical point.  Second row: The average number
  $N_s(t)$ of active sites conditioned on survival plotted against
  $(\ln t)^2$.  Third row: The logarithm of the spread $R(t)$ defined
  in Eq. (\ref{spread}) plotted against $\ln[t/(\ln t)^{2\alpha-4}]$.
  The solid line with a slope $1/\alpha$ indicates the asymptotic
  behavior predicted by the SDRG approach of the model in the critical
  point.}
\end{figure*}

According to the Harris criterion in Eq. (\ref{criterion}), weak disorder is predicted to be relevant in one dimension if $\alpha>3/2$.
Numerical results for $\alpha=2,3$ with $c=0.5$ are shown in
Figs. \ref{1dfigs}d-\ref{1dfigs}i.  
In these cases, a Griffiths phase can be identified, where the survival probability decays algebraically with the time (with possible logarithmic corrections). The decay exponent varies continuously with $\lambda$, and, at a critical value $\lambda_c$, the time-dependence is compatible with the form given in Eq. (\ref{Pt}).
The critical point is estimated to be at $\lambda_c=2.90(1)$ for
$\alpha=2$ and $\lambda_c=5.00(5)$ for $\alpha=3$.  In this point, the
time-dependence of the average number of active sites in 
surviving trials is compatible with
the square-logarithmic law given in Eq. (\ref{Nst}). 
The predicted behavior of the spread
given in Eq. (\ref{Rt}) fits satisfactorily to the data for
$\alpha=2$ but, for $\alpha=3$, a slight discrepancy can be observed.

\begin{figure*}
\includegraphics[width=5.6cm]{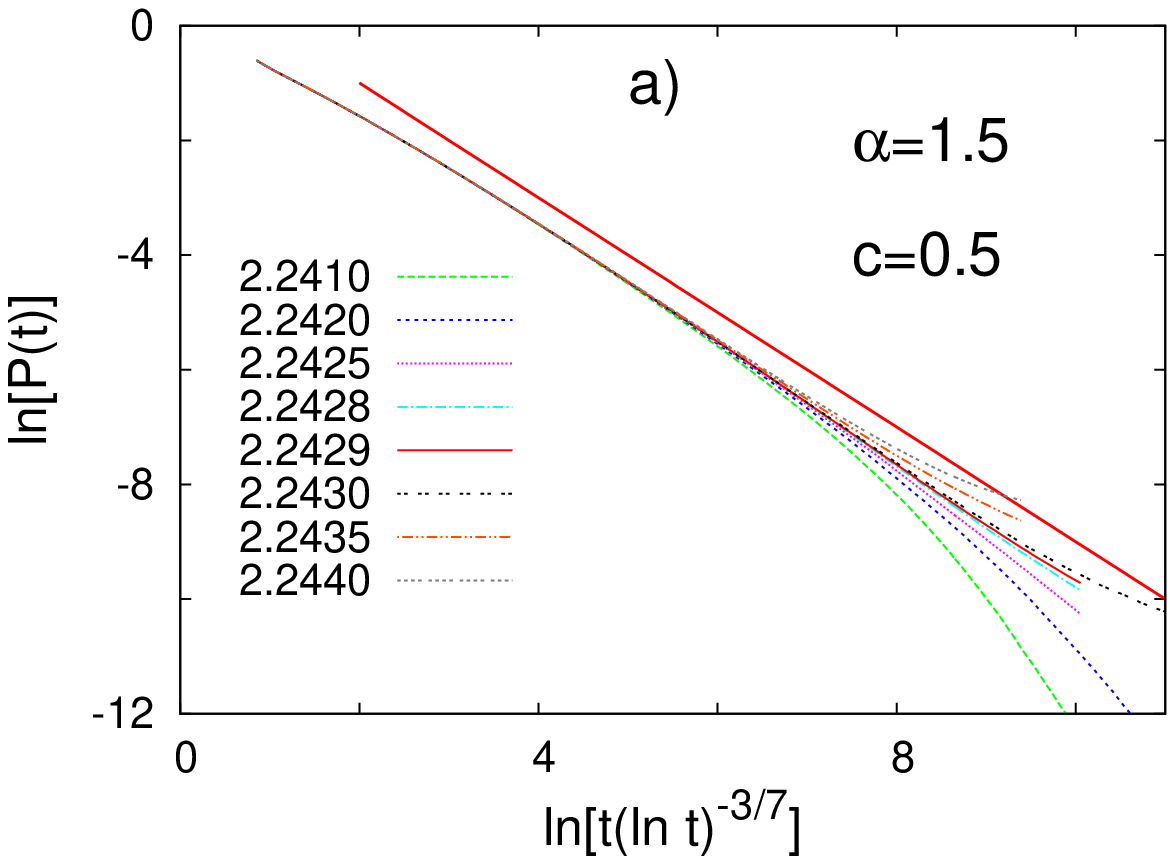}  
\includegraphics[width=5.6cm]{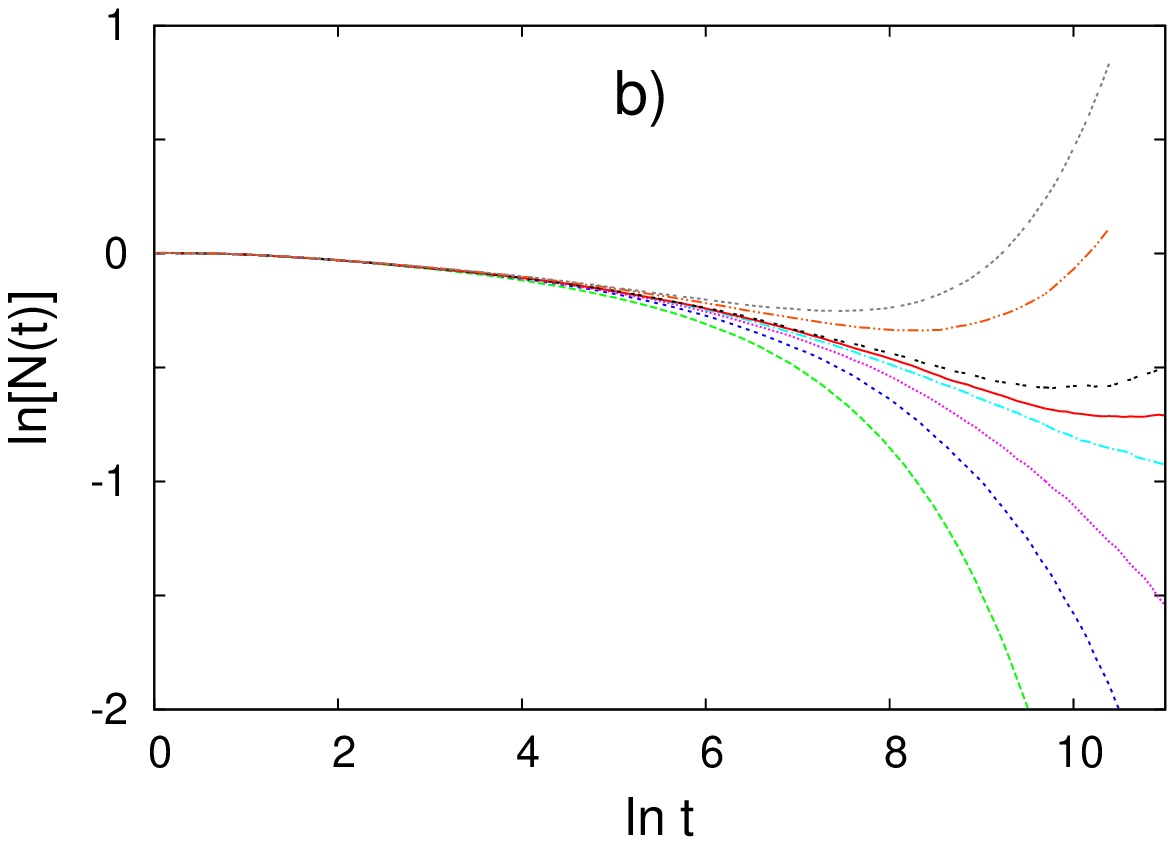}  
\includegraphics[width=5.6cm]{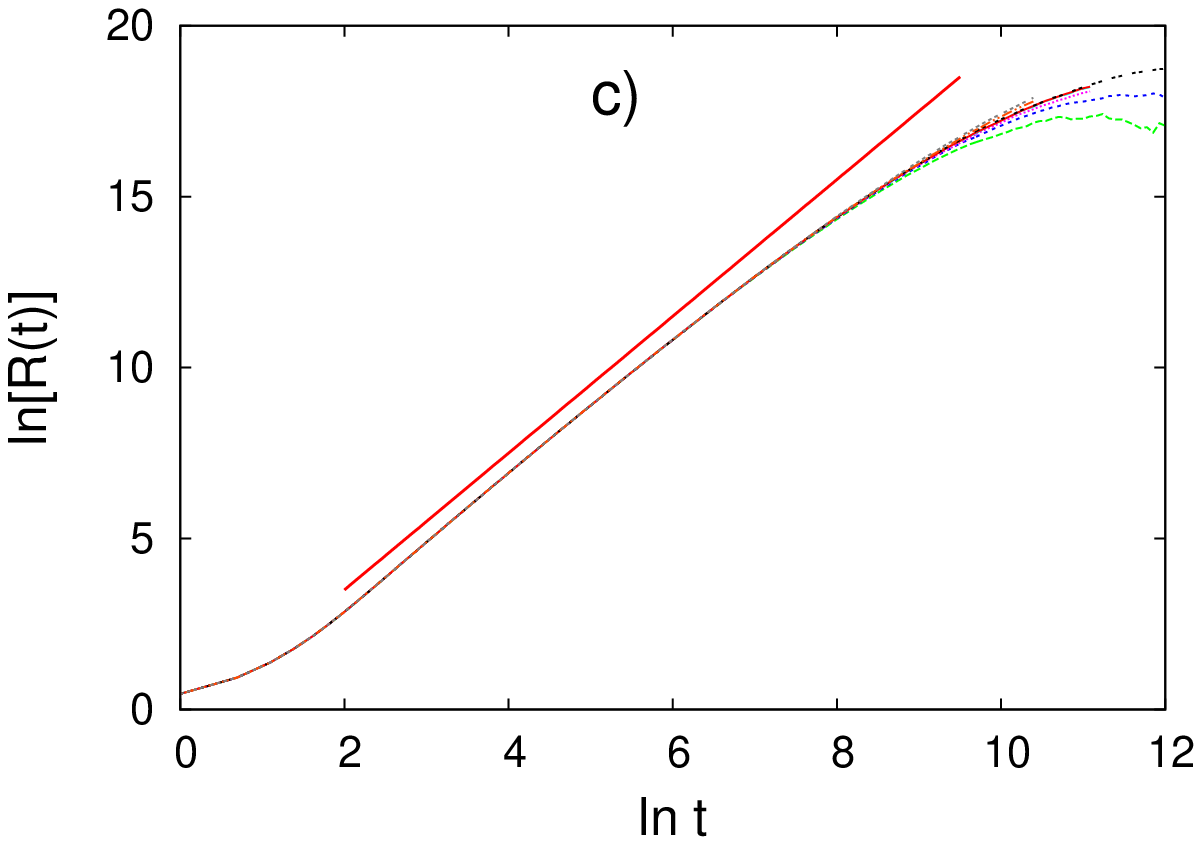}  
\caption{\label{1dmf} (Color online) 
a) The logarithm of the average survival probability $P(t)$
plotted against $\ln[t(\ln t)^{-3/7}]$. The solid line has a slope
$-1$. The data has been obtained by numerical simulations of the
one-dimensional model with $\alpha=3/2$ and $c=0.5$ for different values of the control parameter $\lambda$.  
b) The logarithm of the average number of
active sites $N(t)$ plotted against $\ln t$.  c) The logarithm of the spread $R(t)$ plotted against $\ln t$. The solid line
has a slope $1/\sigma=2$.  
}
\end{figure*}

Simulations with the marginal decay exponent $\alpha=3/2$ and for a 
dilution $c=0.8$ have shown a behavior similar to that found
for $\alpha>3/2$, as can be seen in Figs. \ref{1dfigs}a-\ref{1dfigs}c.  
For a weaker dilution, $c=0.5$, however, the critical behavior 
rather seems to be compatible with that of the clean LRCP, as shown in 
Figs \ref{1dmf}a-\ref{1dmf}c. 
Here, at the border of the
mean-field regime, the survival probability decays as $P(t)\sim
t^{-1}(\ln t)^{3/7}$, the exponent $\eta$, characterizing the growth
of the number of active sites, $N(t)\sim t^{\eta}$, is zero, while the
spread increases as $R(t)\sim t^{1/z}$ with $z=\sigma$ \cite{janssen}.
Note that, due to the rapid increase of the spread ($R(t)\sim t^{2}$
for $\alpha=3/2$), the times for which finite-size effects are
negligible are much shorter than in the case of the strong-disorder
scenario.  From measurements in this limited range of time, it cannot
be decided whether this is the true asymptotic behavior or a crossover
to the strong-disorder fixed point occurs at larger time scales.

\section{Results for the two-dimensional model}
\label{2d}

\subsection{Relation to the one-dimensional model}
Before presenting Monte Carlo results of the two-dimensional model,
let us sketch a simple argument for the scaling behavior of the
quantities of interest in higher dimensions.  Let us start with the
model with a decay exponent $\alpha$, on a $d$-dimensional hypercubic
lattice of (finite but large) size $L$, which is assumed to be an
integer power of $2$. Let us divide the hypercube into $2^d$ smaller
hypercubes and arrange them in chain (in an arbitrary order). Then,
iterate this step for the smaller hypercubes until we arrive at a
linear chain of length $L^d$. After this procedure, almost all
distances $\ell$ between spins in the original hypercube will be in
the order of $O(\ell^d$). So, we roughly obtain in this way a
one-dimensional model with a reduced decay exponent $\alpha/d$.  As a
first guess for the scaling behavior in $d$-dimensions, we take the
formulas obtained in one dimension and replace the quantities 
having a length dimension $\ell$ with $\ell^d$ and $\alpha$ with $\alpha/d$
in them\cite{note}.  This results in 
\beqn P(t)&\sim& [t(\ln
  t)^4]^{-\frac{d}{\alpha}} \label{Ptd} 
\\ R(t) &\sim& [t(\ln
  t)^{4-2\alpha/d}]^{\frac{1}{\alpha}} 
  \label{Rtd}.  
  \eeqn
Which is expected to be correct for the powers of algebraic factors but not those of logarithmic
corrections. Indeed the relations in Eq.(\ref{Ptd}) and Eq.(\ref{Rtd}) 
are in agreement
with Eq.(\ref{rhot}) and Eq.(\ref{z}), respectively, at least up to logarithmic corrections.
The scaling form of $N_s$ in Eq.(\ref{Nst}), which is purely logarithmic, 
is expected to be so for $d>1$, as well: 
\be
N_s(t)\sim (\ln t)^{\chi},
\label{Nsd}
\ee
with a possibly different power $\chi \ge 2$. 

\subsection{SDRG analysis}

In two dimensions, the SDRG method can only be implemented numerically, here we
refer to the SDRG studies of the short-range RTIM in two and higher dimensions.
In order to have a more efficient numerical algorithm,  
we have used the maximum rule and, in the renormalization rule 
in Eq.(\ref{mu_rule}), we omitted the factor $2$. In the one-dimensional
case, these simplifications are found not to modify the critical properties of the system
and the same type of irrelevance is expected to hold in two dimensions, too. For this
problem we have used the algorithm, which has been developed by us for the RTIM and used recently
to analyze the critical properties of the long-range model in one dimension. In the two dimensional
case we have renormalized finite samples of linear size up to $L=64$. The number of samples were
typically $160000$ (at least $4000$ for the largest size). The parameters of the model were chosen uniformly from the intervals
$\Lambda_{ij}\in(0,1]$ and $\mu_i\in(0,\mu]$, with a control parameter, $\theta=\ln(\mu)$.
We have fixed the decay exponent to $\alpha=3$, in which case the critical point is found
at $\theta_c=2.42(5)$ from the analysis of the distributions of the sample-dependent critical points.
(For details of the method we refer the reader to Ref.\cite{jki}).
At the critical point, we have calculated two quantities: 
the average mass of the last remaining
cluster $m(L)=L Q(L)$, see in Eq.(\ref{rhoL}) and the characteristic time scale $\tau(L)$ defined as
$\tau=1/\tilde{\mu}$, where $\tilde{\mu}$ is the last decimated
parameter in a finite sample.

\begin{figure}[ht]
\includegraphics[width=8cm]{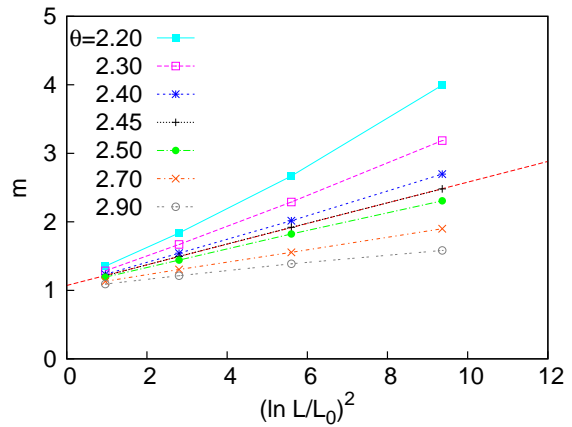}  
\caption{\label{fig_rg_m} (Color online) The average mass of the last
  decimated cluster plotted against $(\ln L/L_0)^2$ with $L_0=3$.
The data has been
  obtained by numerical renormalization of the two-dimensional model
  with decay exponent $\alpha=3$, for different values of the control
  parameter $\Theta$.  }
\end{figure}

The numerical results indicate that $m(L)$ has a slower-than-algebraic 
dependence, which can be written
in analogy with the one-dimensional result in Eq.(\ref{rhoL}) as $m(L) \sim [\ln( L/L_0)]^{\chi}$. Precise determination of $\chi$ from the existing
numerical results is difficult, since it is sensitive
to the value of the reference length, $L_0$.
The data in Fig.\ref{fig_rg_m} are compatible with $\chi=2$ with $L_0=3$, but a similar fit is obtained
with $\chi=3$ if we choose $L_0=0.5$ instead.

\begin{figure}[ht]
\includegraphics[width=8cm]{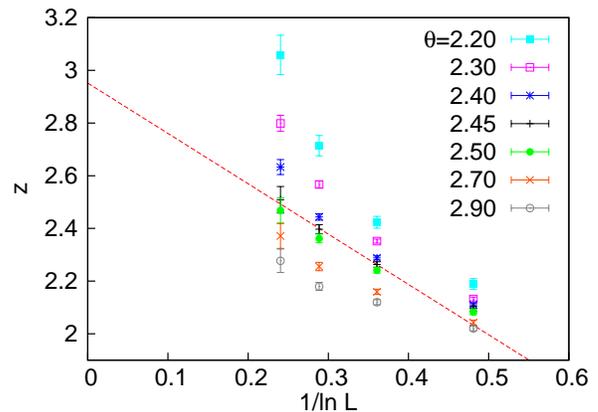} 
\caption{\label{fig_rg_z} (Color online) 
Effective dynamical exponents obtained by two-point fits using Eq. (\ref{tau_L}) as a function of the system size $L$. The straight line is a fit to the data obtained for $\Theta=2.45$. 
}
\end{figure}

Calculating the average logarithmic time scale, 
$\overline{\ln\tilde{\tau}}$,
estimates of an effective, size-dependent dynamical exponent,
$z(L)$, has been obtained from two-point fits of the relation
\be
\overline{\ln\tilde{\tau}}=z\ln L+\mathrm{const}.
\label{tau_L}
\ee
The extrapolation of $z(L)$ to infinite system size, as 
shown in Fig. \ref{fig_rg_z}, is
compatible with the expectation $z_c=\alpha$. 

\subsection{Numerical simulations}
\label{simulation2d}

We have performed Monte Carlo simulations of the two-dimensional model
on diluted lattices of linear size $L=40000$. 
In case of infection events, the target sites have been chosen as described in Ref. \cite{linder}.
For each set of parameters ($\alpha$,$c$,$\lambda$), seed simulations
have been carried out in $100-1000$ randomly diluted lattices, for
$160000$ different starting positions in each sample, and averages of
dynamical quantities have been calculated.

\begin{figure*}
\includegraphics[width=5.6cm]{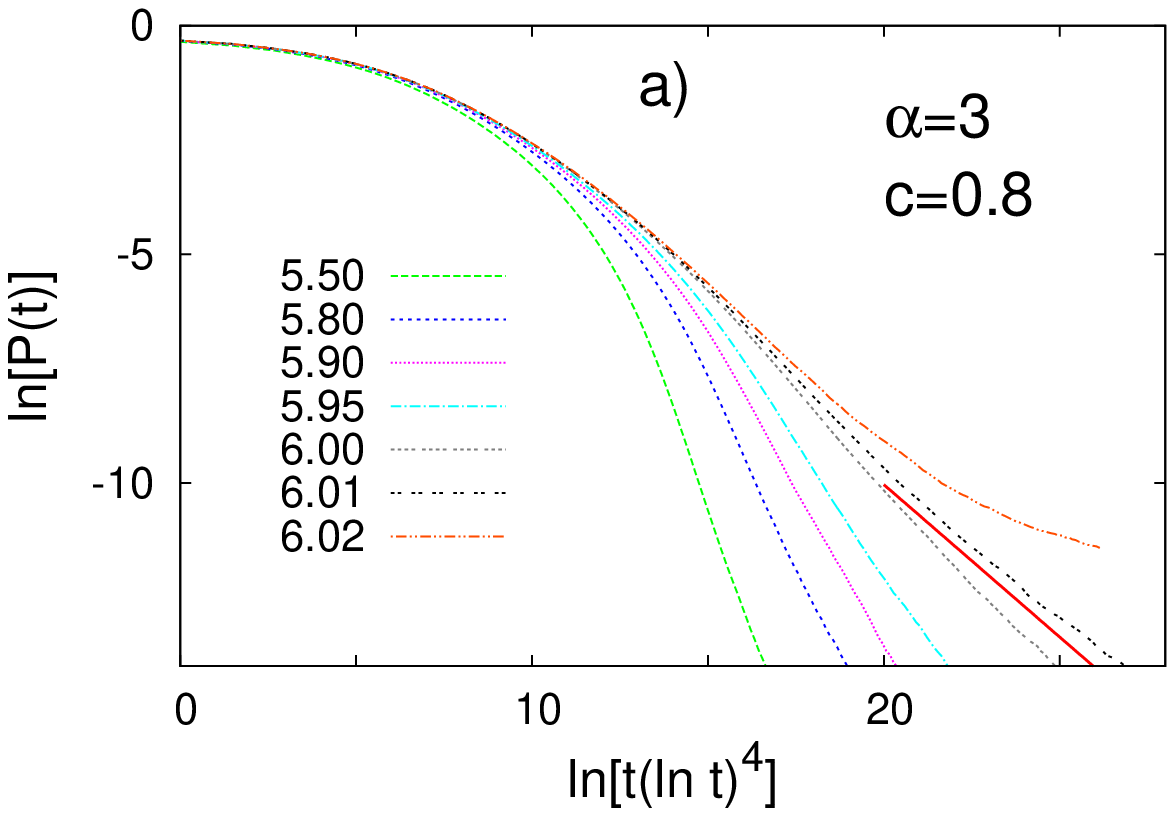}  
\includegraphics[width=5.6cm]{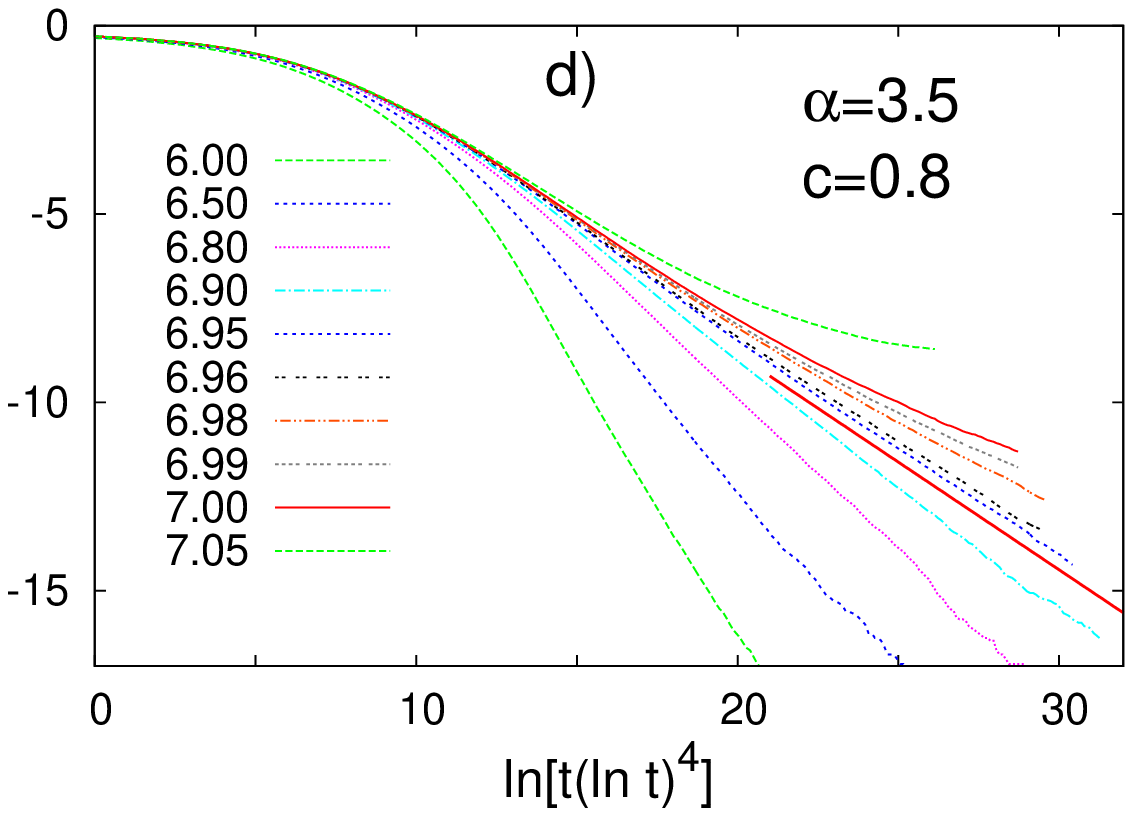}  
\includegraphics[width=5.6cm]{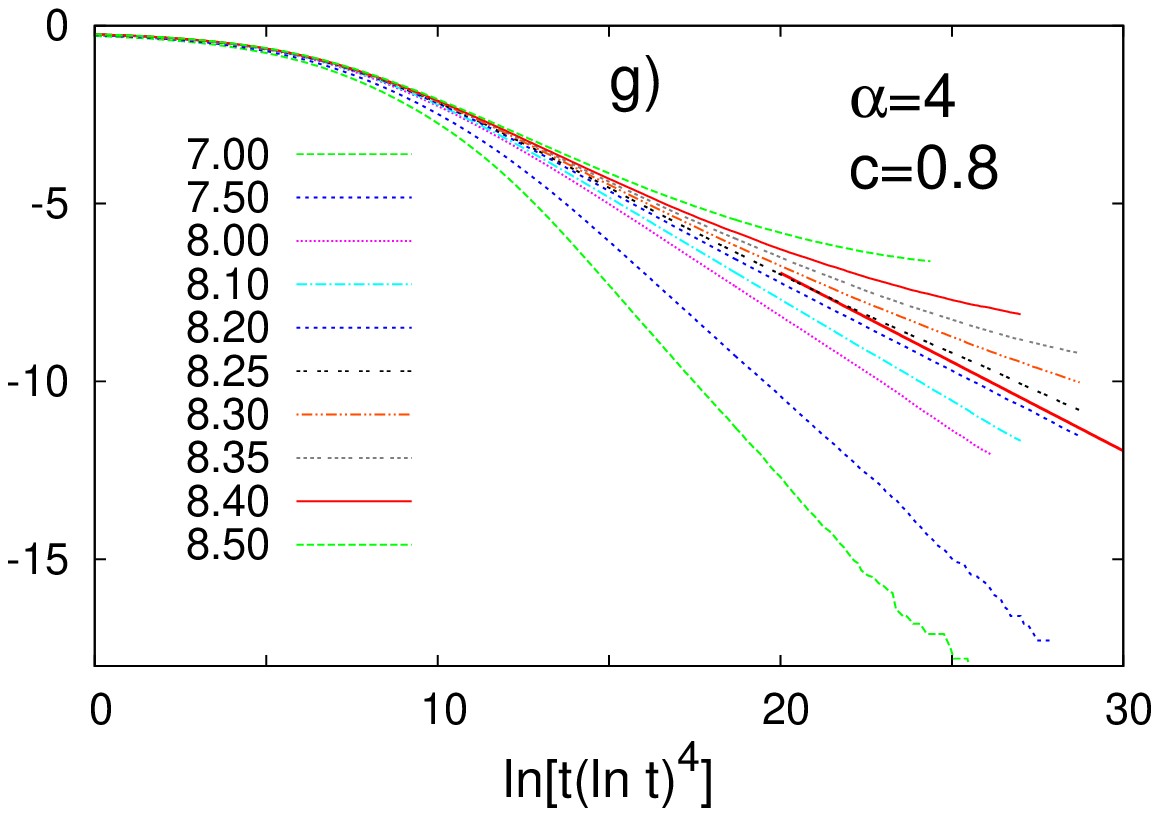}  
\includegraphics[width=5.6cm]{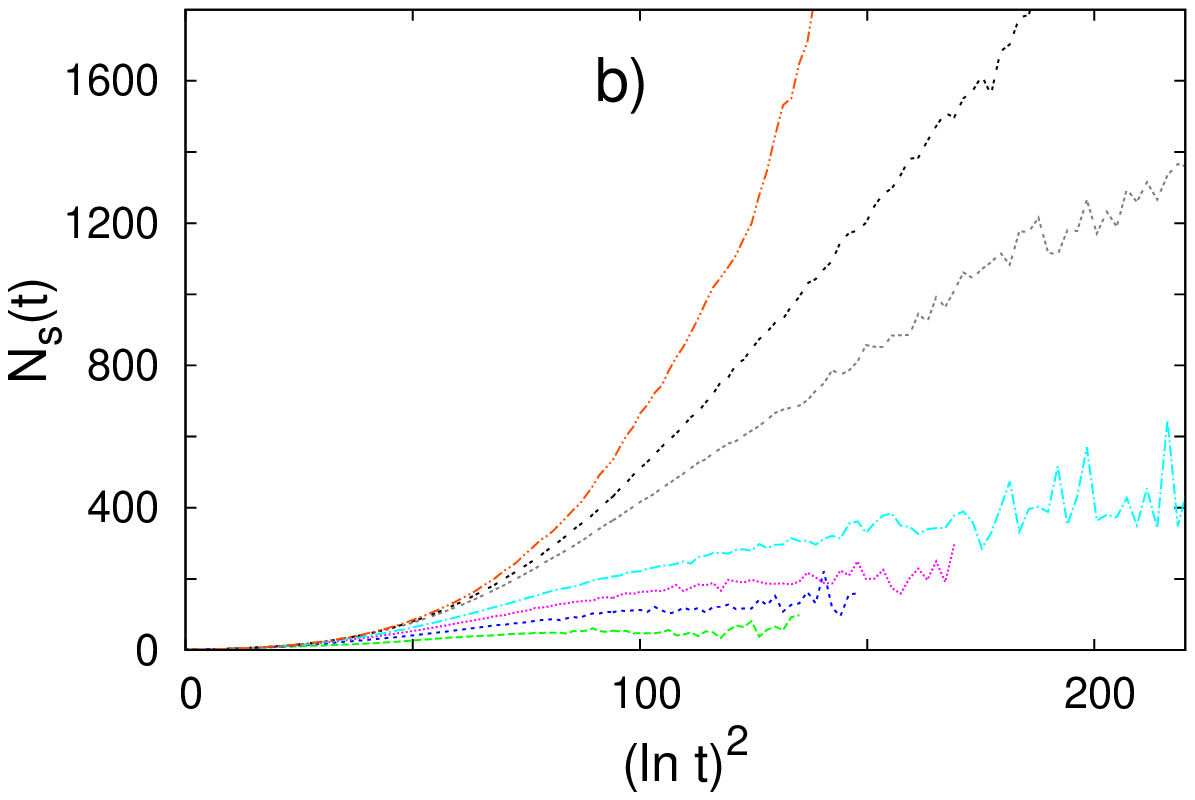}  
\includegraphics[width=5.6cm]{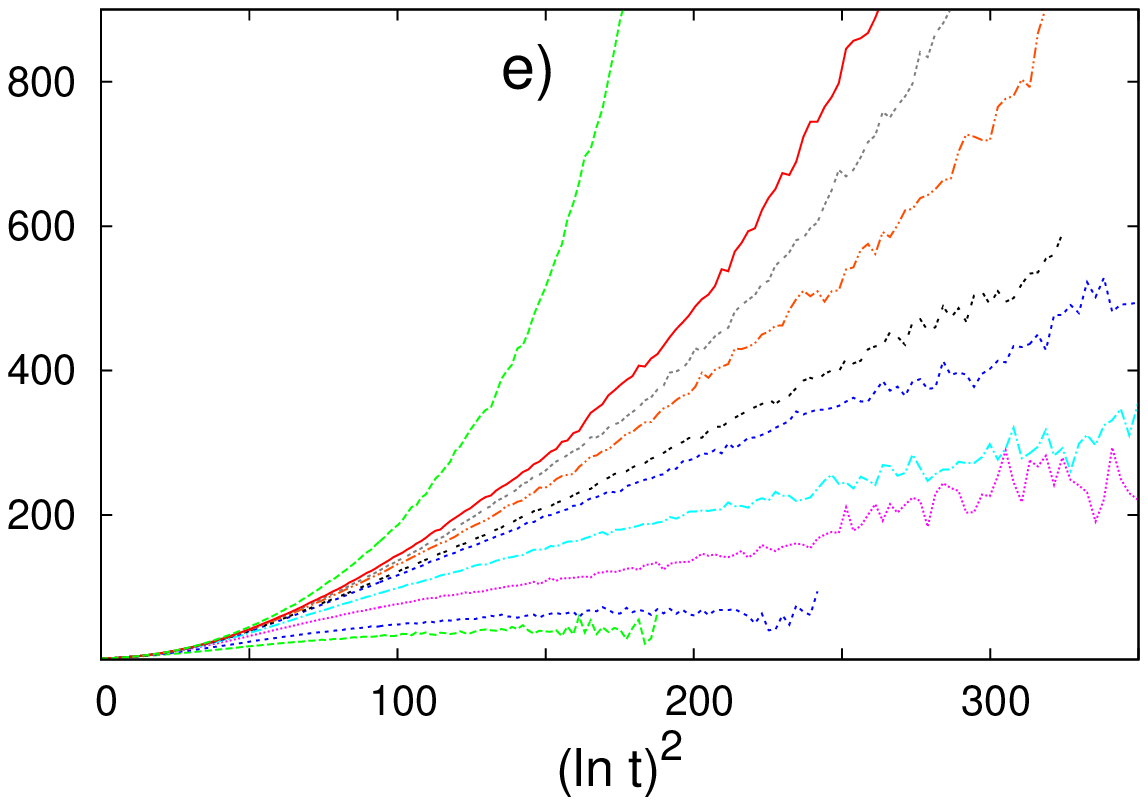}  
\includegraphics[width=5.6cm]{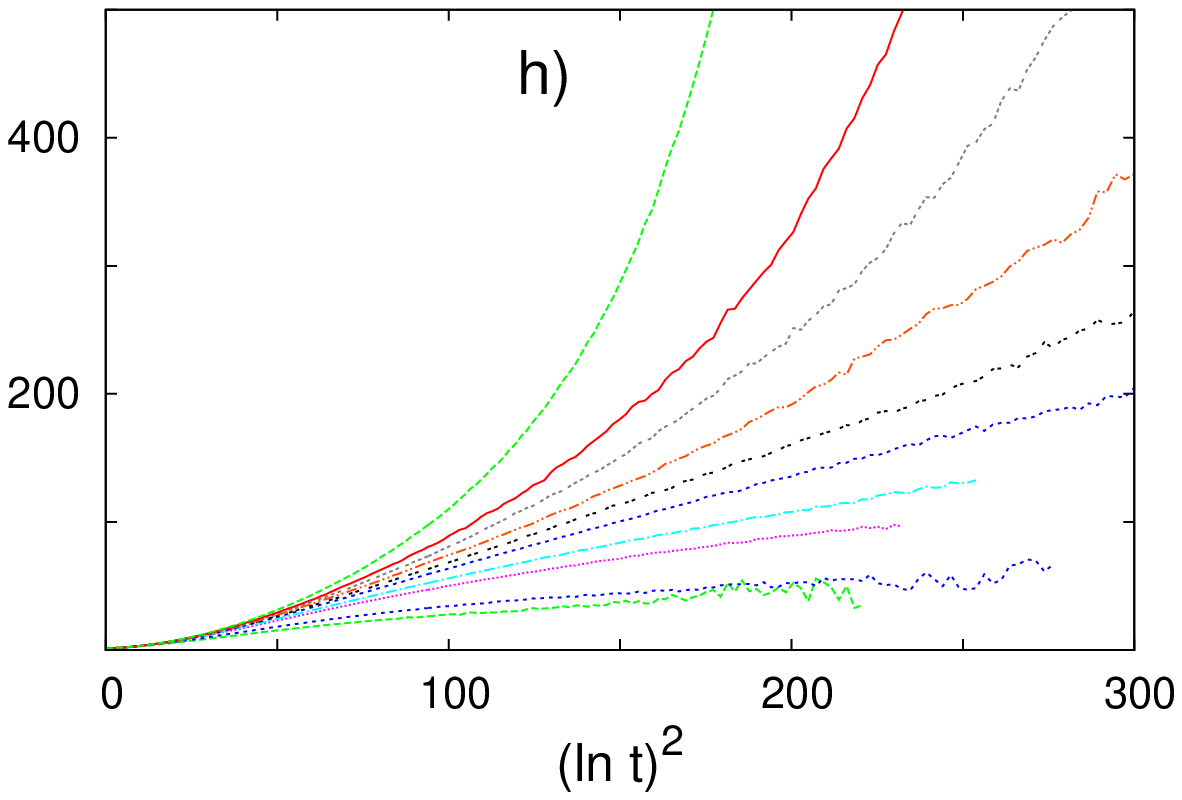}  
\includegraphics[width=5.6cm]{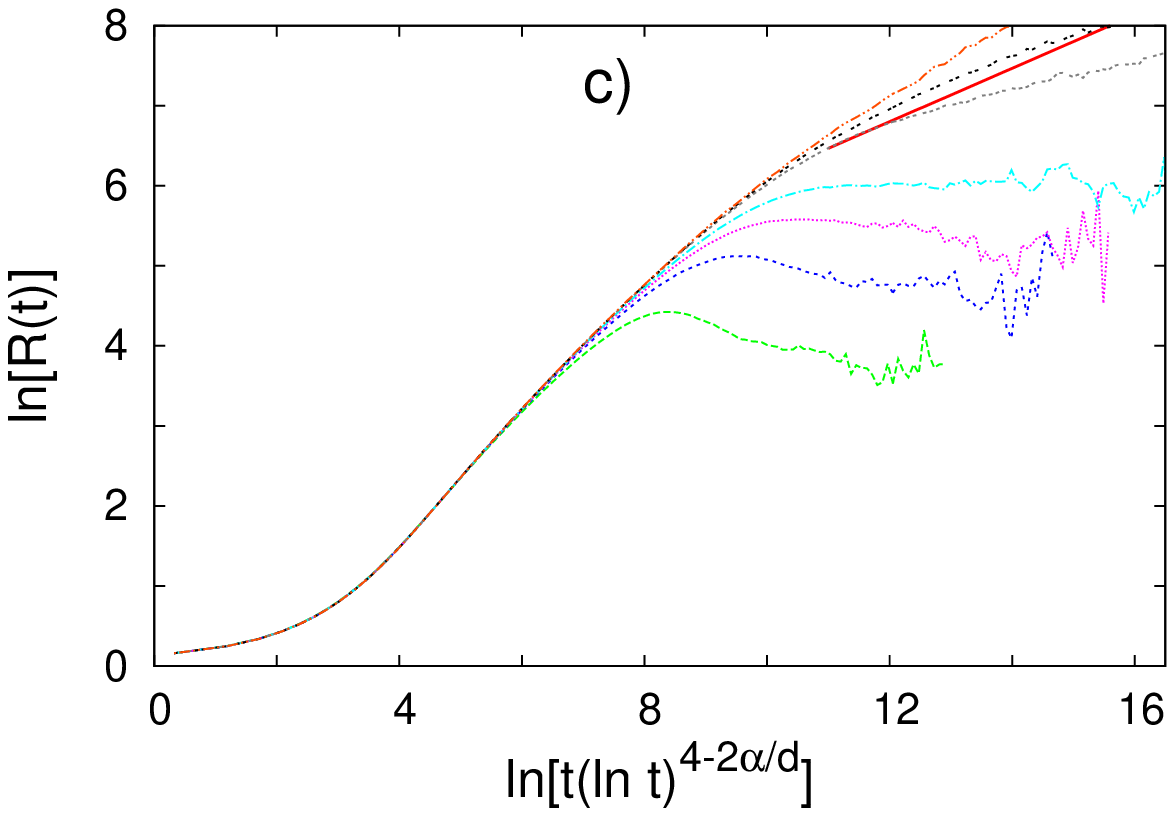}  
\includegraphics[width=5.6cm]{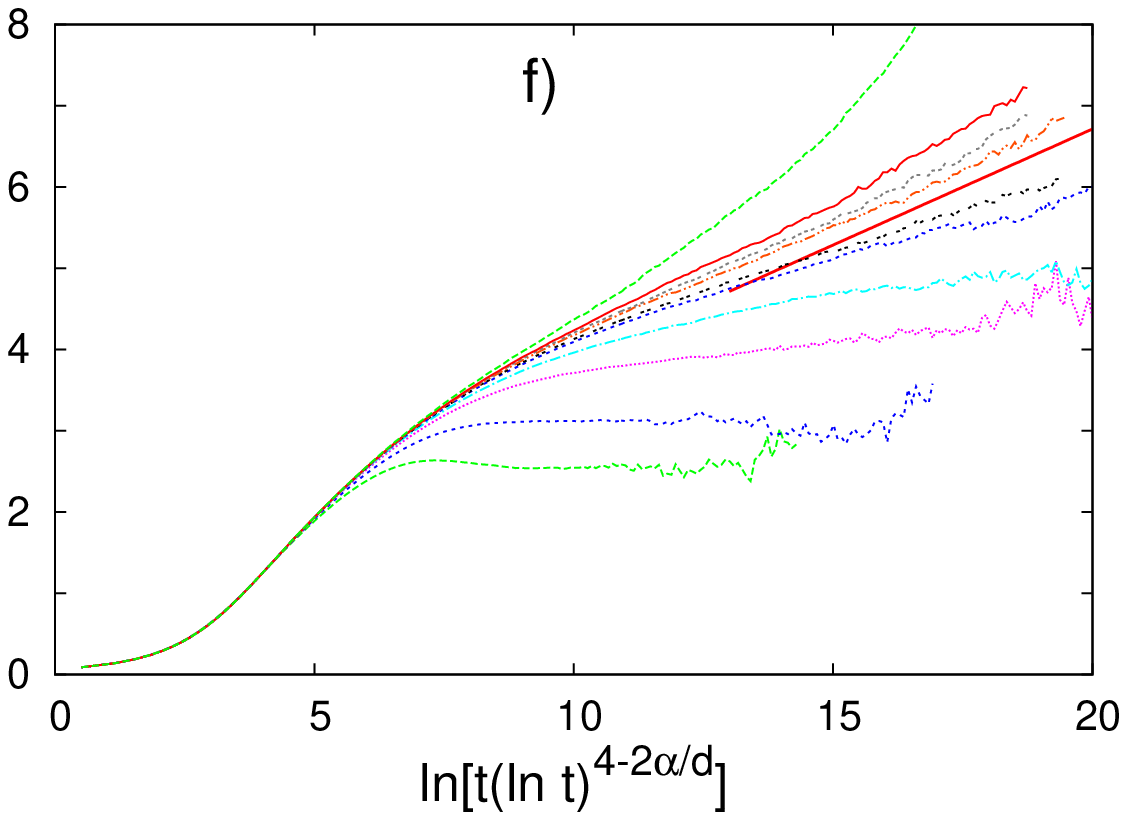}  
\includegraphics[width=5.6cm]{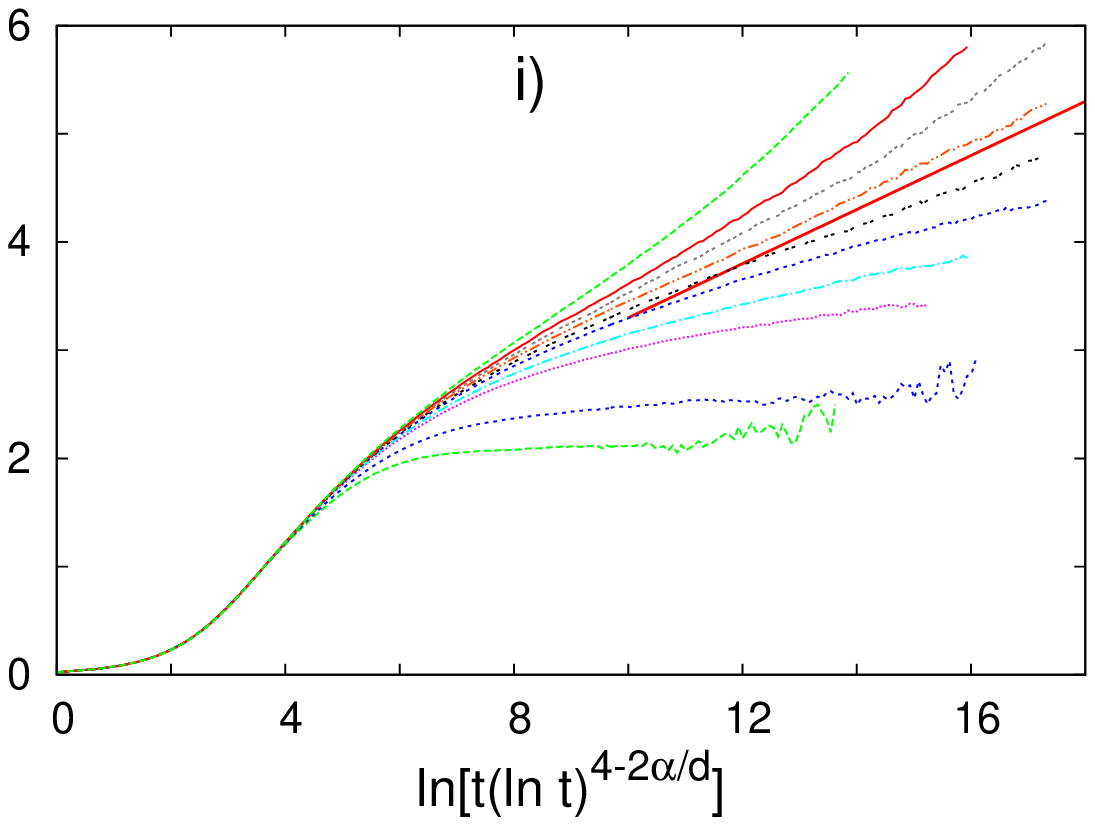}  
\caption{\label{2dfigs} (Color online) First row: The logarithm of the
  average survival probability $P(t)$ plotted against $\ln[t(\ln
    t)^4]$. The solid line has a slope $-d/\alpha$. The data has been
  obtained by numerical simulations of the two-dimensional model for
  different values of the control parameter $\lambda$.  Second row:
  The average number $N_s(t)$ of active sites conditioned on survival
  plotted against $(\ln t)^2$.  Third row: The logarithm of the spread $R(t)$
  defined in Eq. (\ref{spread}) plotted against $\ln[t/(\ln
    t)^{2\alpha/d-4}]$. The solid line has a slope $1/\alpha$.  }
\end{figure*}

In two dimensions, the Harris criterion in Eq. (\ref{criterion}) predicts weak disorder to be relevant if $\alpha>3$.
In this domain, we have
detected a Griffiths phase and obtained that the critical behavior is
satisfactorily compatible with the predictions of the SDRG method extended
to $d>1$, see Eqs. (\ref{Ptd}-\ref{Nsd}), apart from a poor agreement
in the case of the time-dependence of the spread.  
The exponent $\chi$ describing the increase of the average number of active sites in surviving trials in Eq. (\ref{Nsd}) is found to be compatible with the value $\chi=2$ characterizing the one-dimensional model. 
As an illustration,
we show results obtained for $\alpha=3.5$ and $c=0.8$ in
Figs. \ref{2dfigs}d-\ref{2dfigs}f. The critical point is estimated
to be at $\lambda_c=6.95(3)$.  Similar conclusions
have been obtained for $\alpha=4$, $c=0.8$, see Figs. \ref{2dfigs}g-\ref{2dfigs}i, and for $\alpha=3.5$, $c=0.5$ (not shown).

\begin{figure*}
\includegraphics[width=5.6cm]{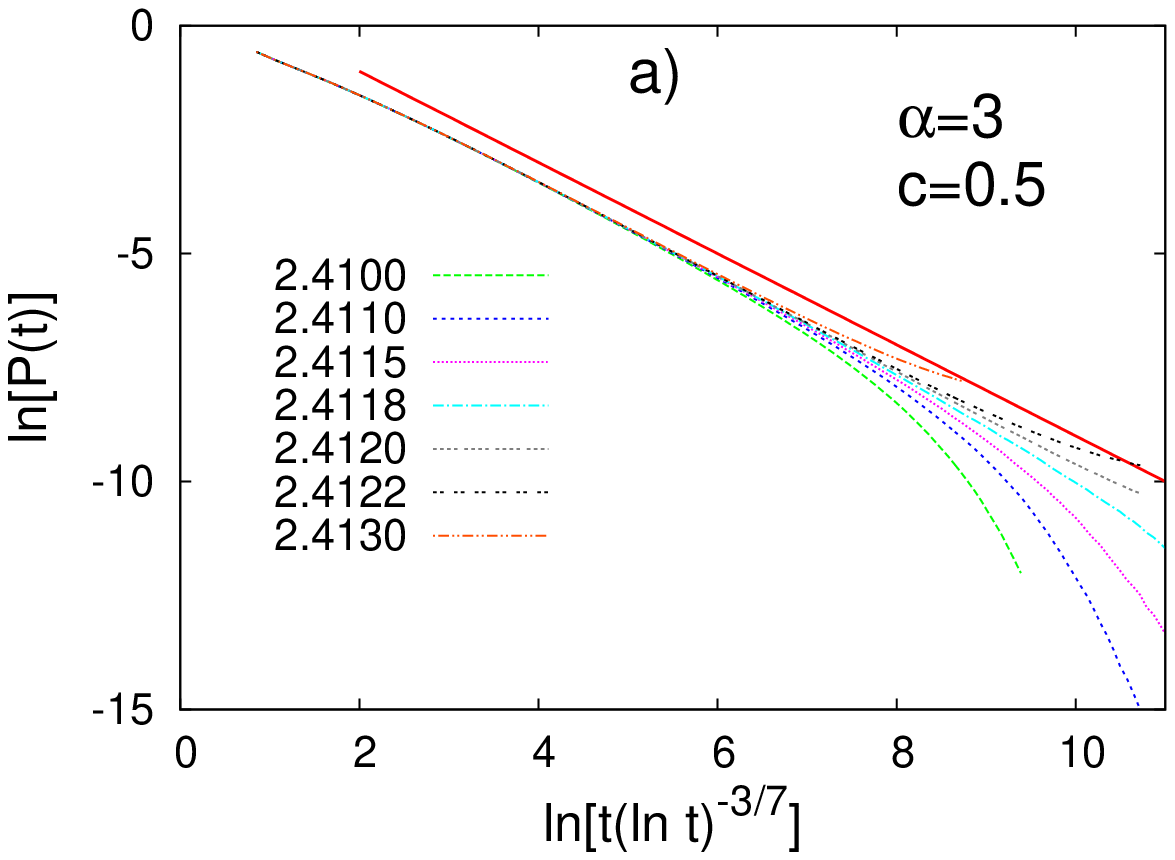}  
\includegraphics[width=5.6cm]{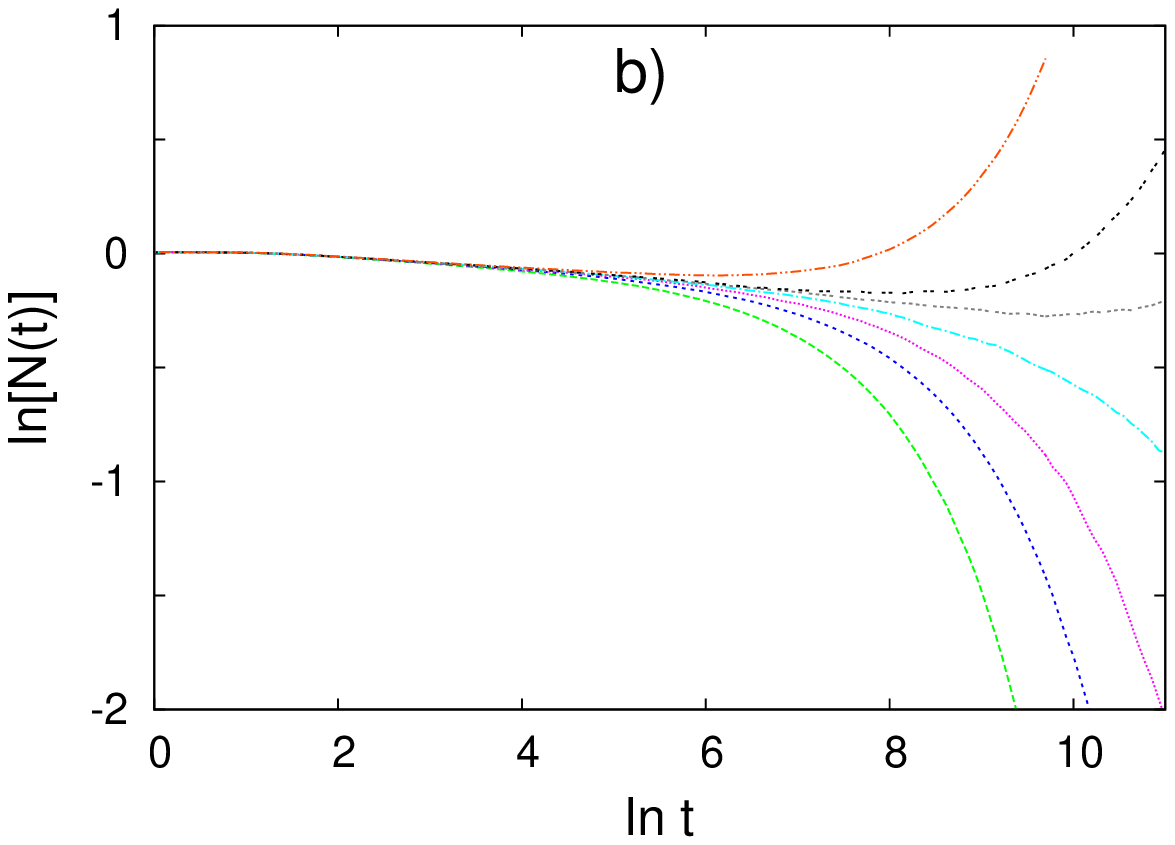}  
\includegraphics[width=5.6cm]{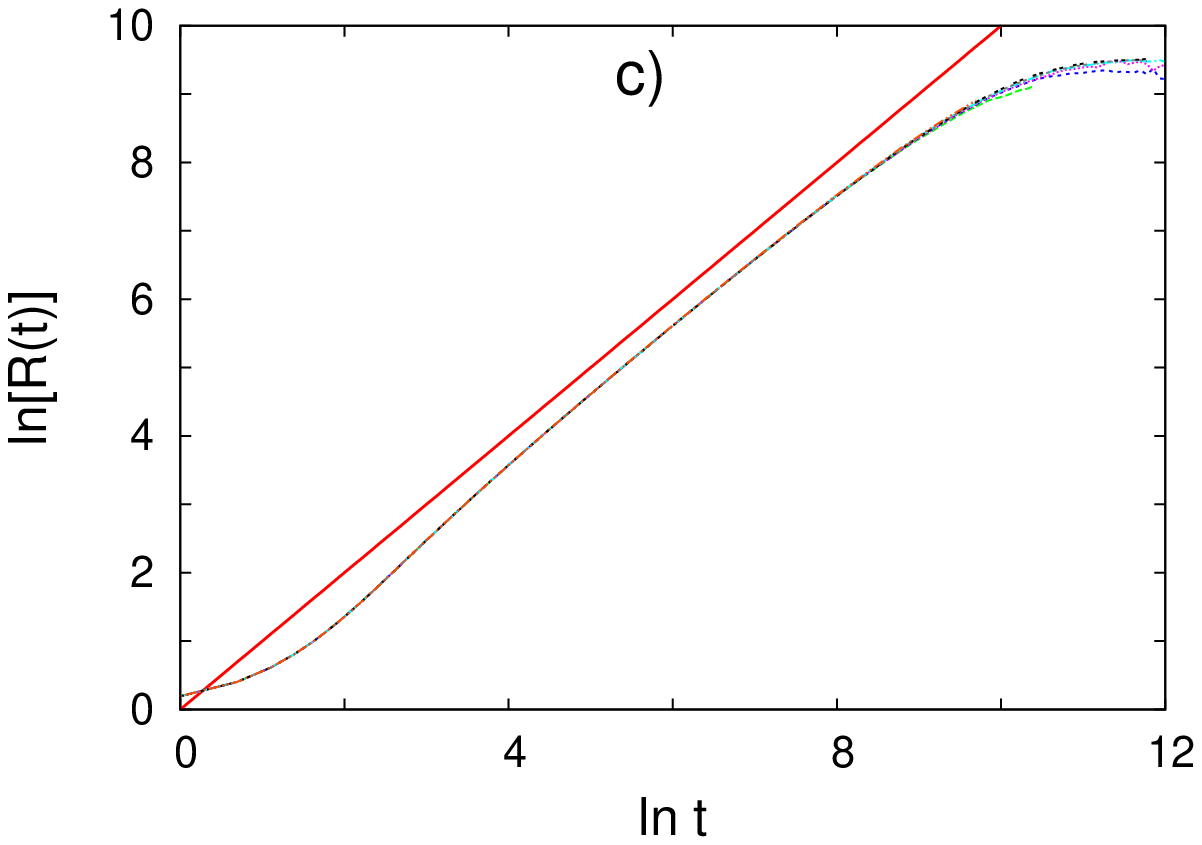}  
\caption{\label{2dmf} (Color online) 
a) The logarithm of the average survival probability $P(t)$
plotted against $\ln[t(\ln t)^{-3/7}]$. The solid line has a slope
$-1$. The data has been obtained by numerical simulations of the
two-dimensional model with $\alpha=3$ and $c=0.5$ 
for different values of the control parameter
$\lambda$. b) The logarithm of the average number of
active sites $N(t)$ plotted against $\ln t$. c) The
logarithm of the spread $R(t)$ plotted against $\ln t$. The solid line
has a slope $1/\sigma=1$.  }
\end{figure*}

For the decay exponent $\alpha=3$, at which, according to the Harris criterion, weak disorder is marginal, the strong-disorder
fixed point still seems to describe the critical behavior of the model 
for a strong enough dilution, as shown for $c=0.8$ in Figs. \ref{2dfigs}a-\ref{2dfigs}c. The critical point is estimated to be at $\lambda_c=6.01(1)$.
For a weaker dilution, however, as it is illustrated by the data in
Figs. \ref{2dmf}a-\ref{2dmf}c obtained for $c=0.5$, no Griffiths effects can be observed and 
the critical behavior seems to agree with that of the clean LRCP, which is described by the mean-field theory, rather than with the strong-disorder scenario. 
These numerical results thus suggest that, fixing the decay exponent to its marginal value $\alpha=3$, the disorder changes from irrelevant to relevant as its strength is increased.  
Note that a similar scenario has been found numerically for the disordered, short-range CP in Ref. \cite{hiv,nft}; see, however, the conflicting results in Ref. \cite{vd,hoyos}. 
We stress again that the time
interval in which the model is free from finite-size effects, see
Fig. \ref{2dmf}c, is rather short, and in this regime 
no definite conclusions can
be drawn concerning the asymptotic behavior from the numerical data.

We have simulated the model with $\alpha=2.5$ and $c=0.8$, as well,
which belongs to the domain where weak disorder is predicted to 
be irrelevant by the Harris criterion. 
Here, mean-field critical behavior is found (not shown), and,  
within the very short time scale,
where the system is free from finite-size effects, no indications of the
existence of a Griffiths phase has been seen.
It is interesting to compare these observations with the behavior of  
the disordered, short-range contact process above its
upper critical dimension $d_c=4$, where weak disorder is irrelevant, 
just as in our case.   
According to a recent conjecture, which is supported by simulation results for $d=5$ \cite{5d}, the critical behavior is of mean-field type but a Griffiths phase still shows up, where the dynamical exponent varies with the control parameter and saturates to the mean-field value.      
The numerical results obtained for our model fit only partially to this scenario, as Griffiths effects could not be observed in our case.  

\section{Discussion}
\label{discussion}
In this paper, we have studied a long-range contact process in
a random environment. This model is realized in different situations, when the agents transmitting the disease can be insects, human beings or spores. 
Depending on the relative
strengths of the infection and the recovery rates, this model exhibits an active (endemic) phase with a finite fraction of infected sites in the steady state and an inactive one, where all sites are healthy. 
The properties of the nonequilibrium phase transition
in the system are studied by different methods in one and two spatial dimensions.

Analytical results are obtained in one dimension by a variant of the SDRG method, which is called the primary scheme. 
The critical point is found
to be controlled by a so called strong-disorder fixed point, in which the dynamical exponent
is finite and given by $z_c=\alpha$ in any dimension. 
The average number of active sites in surviving samples is found to increase as $N_s(t) \sim (\ln t)^{\chi}$, with
$\chi=2$ in $d=1$, thus, the set of active sites has a formally zero fractal dimension.  
In the nearby inactive Griffiths phase,
the decay of the average density and the growth of the average spatial extent of the set of active sites still follow power-laws, however the dynamical exponent is $z<\alpha$ and continuously
depends on the distance from the transition point.
The theoretical predictions obtained by the approximative SDRG method and phenomenological reasoning have been confronted with results of large-scale 
Monte Carlo simulations. 
We have found satisfactory agreement in the non-mean-field regime of the clean model, $\alpha>\frac{3}{2}d$, where Harris criterion predicts relevance of weak disorder. 
At the boundary of this regime, $\alpha=\frac{3}{2}d$, the numerical data are compatible with the strong-disorder scenario only for a strong enough 
initial disorder; 
otherwise, as well as for $\alpha<\frac{3}{2}d$, they rather seem to follow the mean-field theory of the clean model.  
We emphasize, however, that, for judging the long-time asymptotic behavior in the mean-field regime, the present numerical results are not decisive, and a crossover to a strong-disorder fixed point at large scales, even for a weak disorder, cannot be excluded. 
We mention that, even in the simplest case, 
i.e. in the one-dimensional, short-range CP, although large-scale simulations suggest a positive answer \cite{vd},
it is an unresolved question whether the IDFP is attractive for any weak disorder\cite{crossover}. The situation observed in our model for $\alpha\le\frac{3}{2}d$ is similar to that of the short-range CP at and above the upper critical dimension $d_c=4$, where, although the SDRG approach predicts an IDFP in any dimensions for a strong enough disorder \cite{kovacs}, numerical simulations indicate a mean-field critical behavior \cite{5d}. How the nature of the phase transition changes with the strength of the disorder in this case is a puzzling question.

It is worth comparing the behavior of our model to that of the contact process on static (i.e. time-independent) random networks embedded in a d-dimensional space and having long-range links between remote sites with an algebraically decaying connection probability, $p(l_{ij})\sim l_{ij}^{-s}$ \cite{munoz,jk}. 
This model can be regarded as a ``quenched'' variant of the LRCP, where the randomly drawn LR links induce quenched ``topological'' disorder. 
According to a numerical SDRG analysis of the one-dimensional model \cite{jk},
for $s\ge 2$, the critical behavior is controlled by an IDFP, which is identical to that of the short-range disordered CP for $s>2$.  
For $s<2$, Monte Carlo simulations of the model show that the critical dynamics 
is described by power-laws and no signs of Griffiths effects can be observed. 
An apparent difference between the two models is that the quenched model displays short-range behavior for a sufficiently large decay exponent ($s>2$), while this is not the case for the disordered LRCP.

The results presented in this work are similar to that obtained very recently for the long-range RTIM. Indeed, the SDRG
decimation steps differ only at one point in the two problems, which has been shown to be irrelevant
in one dimension, see the Appendix. The same conclusion is expected to hold in higher dimensions, too,
which is demonstrated in Sec.\ref{2d} for the two-dimensional case. 
For the RTIM, analysis of the primary model
on the paramagnetic side of the critical point, which corresponds to the inactive phase of our model, leads to a divergence of the correlation length $\xi$ of the form
\be
\xi \sim \exp(A/|\theta-\theta_c|)\;,
\label{xi}
\ee
which resembles the singularity present in the Kosterlitz-Thouless 
transition. We expect that Eq.(\ref{xi}) holds
for the LRCP, as well. However, a numerical verification of this conjecture is very difficult.

\begin{acknowledgments}
This work was supported by the National Research Fund under grant
No. K109577; by the J\'anos Bolyai Research
Scholarship of the Hungarian Academy of Sciences (RJ), and partially
supported by the European Union and the European Social Fund through
project FuturICT.hu (grant no.: TAMOP-4.2.2.C-11/1/KONV-2012-0013).
The research of IAK was supported by the European Union and the State
of Hungary, co-financed by the European Social Fund in the framework
of T\'AMOP 4.2.4. A/2-11-1-2012-0001 'National Excellence Program'.
\end{acknowledgments}

\appendix
\section{The primary SDRG scheme in one dimension}

Let us restrict ourselves to one dimension and assume that
$\Lambda_{ij}=1$ in Eq. (\ref{lambda}), i.e. the activation rates are
non-random, but the deactivation rates are still random.  This model
is, according to numerical investigations, in the same universality
class as the one with random activation rates but it is simpler to
treat analytically. Analyzing the SDRG procedure close to the fixed point in the inactive phase and in the critical point,
we have a few observations, which can be used to simplify the SDRG scheme. These observations are
analogous to those found for the one-dimensional RTIM\cite{jki}. First, almost always cluster eliminations occur; 
second, after decimating a cluster, the maximum rule leads almost always to $\tilde{\lambda}_{jk}=\lambda_{jk}$;
third, the extensions of (non-decimated) clusters are typically much smaller than the distances between them. 
These lead to a simplified scheme of effectively one-dimensional structure, in which only the activation rates between neighboring clusters are renormalized. 
Introducing the reduced parameters
$\zeta=(\frac{\Omega}{\lambda})^{1/\alpha}-1$ and
$\beta=\frac{1}{\alpha}\ln\frac{\Omega}{\mu}$,
in analogy with the RTIM, the renormalization rules assume the additive forms 
\be
\tilde\zeta=\zeta_{i-1,i}+\zeta_{i,i+1}+1, 
\ee
if a cluster is decimated and 
\be
\tilde\beta=\beta_i+\beta_{i+1}-B 
\ee
if two clusters are unified. 
The difference compared to the SDRG rules of the
long-range RTIM is the appearance of constant $B=\frac{1}{\alpha}\ln 2$.  We
mention that the SDRG scheme with $B= 0$
first arose in the context of a disordered quantum rotor model
\cite{akpr}.  As the logarithmic rate scale
\be
\Gamma\equiv\frac{1}{\alpha}\ln\frac{\Omega_0}{\Omega}\;,
\label{Gamma}
\ee
with
$\Omega_0$ being the initial value of $\Omega$, increases during the
procedure, the distributions $g_{\Gamma}(\beta)$ and
$f_{\Gamma}(\zeta)$ evolve according to the equations
\begin{widetext}
\beqn  
\frac{\partial g_{\Gamma}(\beta)}{\partial\Gamma}=
\frac{\partial g_{\Gamma}(\beta)}{\partial\beta} +  
f_0(\Gamma)\int d\beta_1\int d\beta_2g_{\Gamma}(\beta_1)g_{\Gamma}(\beta_2)
\delta[\beta-\beta_1-\beta_2+B] +
g_{\Gamma}(\beta)[g_0(\Gamma)-f_0(\Gamma)]  \label{gflow} \\
\frac{\partial f_{\Gamma}(\zeta)}{\partial\Gamma}=
(\zeta+1)\frac{\partial f_{\Gamma}(\zeta)}{\partial\zeta} +  
g_0(\Gamma)
\int d\zeta_1\int
d\zeta_2f_{\Gamma}(\zeta_1)f_{\Gamma}(\zeta_2)\delta(\zeta-\zeta_1-\zeta_2-1) +
f_{\Gamma}(\zeta)[f_0(\Gamma)+1-g_0(\Gamma)],
\label{fflow}
\eeqn
\end{widetext}
where $g_0(\Gamma)\equiv g_{\Gamma}(0)$ and $f_0(\Gamma)\equiv f_{\Gamma}(0)$.

In the critical point and below, the variables $\zeta$ will grow
without limits as $\Gamma\to\infty$, therefore the constant in the
delta function in Eq. (\ref{fflow}) can be neglected and the fixed
point distribution will be
$f_{\Gamma}(\zeta)=f_0(\Gamma)e^{-f_0(\Gamma)\zeta}$, where
$f_0(\Gamma)$ satisfies the differential equation 
\be
\frac{df_0(\Gamma)}{d\Gamma}=f_0(\Gamma)[1-g_0(\Gamma)].
\label{f_diff}
\ee
For $B=0$, the fixed point distribution
$g_{\Gamma}(\beta)$ also will be a pure exponential, but, otherwise,
it is no longer the case.  Assuming that $g_0(\Gamma)$ tends to a
finite limit $g_0(\infty)$ as $\Gamma\to\infty$, $f_0(\Gamma)$ will
diverge or vanish if $g_0(\infty)<1$ or $g_0(\infty)>1$, respectively,
as can be seen from Eq. (\ref{f_diff}). Thus, in the critical point,
we must have $g_0(\infty)=1$. Here, let us assume that
$g_0(\Gamma)\simeq 1+\frac{G}{\Gamma^{\gamma}}$ and
$f_0(\Gamma)\simeq\frac{F}{\Gamma^{\phi}}$ for large $\Gamma$, where
$G$, $\gamma$, $F$, and $\phi$ are positive constants.  Substituting
these asymptotic forms into Eq. (\ref{f_diff}), we obtain that
$\gamma=1$ and $G=\phi$.  Although $g_{\Gamma}(\beta)$ is not a pure
exponential, its large-$\beta$ tail is still exponential,
$g_{\Gamma}(\beta)\simeq g_1(\Gamma)e^{-g_2(\Gamma)\beta}$, as one can
check by substituting it into Eq. (\ref{gflow}).  This
gives for the unknown functions $g_1(\Gamma)$ and $g_2(\Gamma)$ the
following differential equations 
\beqn
\frac{dg_1(\Gamma)}{d\Gamma}\simeq
g_1(\Gamma)[g_0(\Gamma)-f_0(\Gamma)-g_2(\Gamma)]
\label{dg1}
\\
\frac{dg_2(\Gamma)}{d\Gamma}\simeq -f_0(\Gamma)g_1(\Gamma)e^{-B g_2(\Gamma)}.
\label{dg2}
\eeqn
Let us assume, that these functions have finite limiting values, $g_1(\Gamma)\to G_1$
and $g_2(\Gamma)\to G_2$, as $\Gamma\to\infty$. 
Then, for large $\Gamma$, Eq. (\ref{dg2}) assumes the form  
\be 
\frac{dg_2(\Gamma)}{d\Gamma}\simeq -FG_1e^{-B G_2}\Gamma^{-\phi}
\equiv - \mathcal{G}\Gamma^{-\phi}.
\label{dg2a}
\ee
The finiteness of $G_2$ then requires $\phi>1$, and the integration of Eq.
(\ref{dg2a}) results in 
\be 
g_2(\Gamma)\simeq G_2 + \mathcal{G}\frac{\Gamma^{1-\phi}}{\phi-1}.
\label{g2}
\ee
The asymptotic form of Eq. (\ref{dg1}) then reads as 
\be
\frac{dg_1(\Gamma)}{d\Gamma}\simeq
G_1\left[1+\frac{\phi}{\Gamma}-\frac{F}{\Gamma^{\phi}}
- G_2 - \mathcal{G}\frac{\Gamma^{1-\phi}}{\phi-1}\right].
\label{dg1a}
\ee 
For the reason that $g_1(\Gamma)$ converges in the limit
$\Gamma\to\infty$, the constant and $O(1/\Gamma)$ terms in the
brackets in Eq. (\ref{dg1a}) must cancel, which yields $G_2=1$,
$\phi=2$ and $\mathcal{G}=2$.  So, we conclude that the leading order
$\Gamma$-dependence of the functions $g_0(\Gamma)$ and $f_0(\Gamma)$
in the critical point are given to be 
\be
g_0(\Gamma)\simeq
1+\frac{2}{\Gamma} \qquad f_0(\Gamma)\simeq\frac{F}{\Gamma^2}.
\label{gf}
\ee
The constant $F$, which takes the value $2$ for $B=
0$, remains unknown in the present case. 

A relationship between the length scale $\ell(\Gamma)=1/n(\Gamma)$,
where $n(\Gamma)$ is the mean number of non-decimated clusters per
unit length of the chain and the rate scale $\Omega$ can be derived by
solving the differential equation
$\frac{dn(\Gamma)}{d\Gamma}=-n(\Gamma)[g_0(\Gamma)+f_0(\Gamma)]$.
Using the asymptotic forms in Eq. (\ref{gf}), one obtains 
\be
\ell\simeq\ell_0
e^{\Gamma}\Gamma^2,
\label{dyn0}
\ee
where $l_0$ is a non-universal constant that depends on the initial distribution
of parameters. 

In order to infer the scaling of the order parameter in the critical
point, a further quantity we need is the probability $Q(\Gamma)$ that
a given site is part of an active (non-decimated) cluster at the scale
$\Gamma$.  This can be obtained through the probability density
$q_{\Gamma}(\beta)$ of the event that, at the scale $\Gamma$, a given
site is part of an active cluster that has a logarithmic deactivation
rate $\beta$. This function evolves according to the equation
\begin{widetext}
\be 
\frac{\partial q_{\Gamma}(\beta)}{\partial\Gamma}=
\frac{\partial q_{\Gamma}(\beta)}{\partial\beta} +  
2f_0(\Gamma)\left[
\int d\beta_1\int
d\beta_2q_{\Gamma}(\beta_1)g_{\Gamma}(\beta_2)\delta[
\beta-\beta_1-\beta_2+B] -
q_{\Gamma}(\beta)\right].
\label{qflow}
\ee
\end{widetext}
Knowing the large-$\beta$ tail of $g_{\Gamma}(\beta)$ only, the best
we can do is to determine that of the function $q_{\Gamma}(\beta)$
from this equation.  For large $\beta$, the solution will have the
form $q_{\Gamma}(\beta)\simeq
[q_1(\Gamma)+q_2(\Gamma)\frac{\beta}{\Gamma}]e^{-g_2(\Gamma)\beta}$,
where $q_1(\Gamma)$ and $q_2(\Gamma)$ are unknown functions.
Substituting these asymptotic forms into Eq. (\ref{qflow}), we obtain
that the above functions must satisfy the differential equations 
\beqn
\frac{dq_1(\Gamma)}{d\Gamma}\simeq
-\left(1+\frac{2}{\Gamma}\right)q_1(\Gamma)+\frac{1}{\Gamma}q_2(\Gamma) +
O[\Gamma^{-2}q_1(\Gamma)] \nonumber
\\ \frac{dq_2(\Gamma)}{d\Gamma}\simeq
\frac{\mathcal{G}}{\Gamma}q_1(\Gamma)-\left(1+\frac{1}{\Gamma}\right)q_2(\Gamma)
+ O[\Gamma^{-2}q_1(\Gamma)]. \nonumber \\
\eeqn
Since $\mathcal{G}=2$,
these equations are identical to those of the model with
$B=0$.  The solutions are
$q_1(\Gamma)\simeq c_1e^{-\Gamma}$, $q_2(\Gamma)\simeq
c_2e^{-\Gamma}$, and assuming that the integral
$Q(\Gamma)=\int_0^{\infty}d\beta q_{\Gamma}(\beta)$ scales with
$\Gamma$ in the same way as the contribution from the large-$\beta$
tail of $q_{\Gamma}(\beta)$, we obtain 
\be
Q(\Gamma)\sim e^{-\Gamma}=\left(\frac{\Omega}{\Omega_0}\right)^{1/\alpha}.
\label{Qgamma}
\ee


\begin{thebibliography}{99}
\bibitem{note}
We note, that this type of rescaling is compatible with the results of the leading $\epsilon$-expansion
of the non-random LRCP, see in Ref.\cite{janssen}.
\end{thebibliography}
\end{document}